\documentclass[a4paper,preprint,12pt]{elsarticle}

\usepackage{amssymb}
\usepackage{amsmath}
\usepackage{subfigure}
\usepackage{multirow}
\usepackage{placeins}
\usepackage{siunitx}
\usepackage{graphicx}
\usepackage{upgreek}

\usepackage{verbatim}

\usepackage{color}
\usepackage[normalem]{ulem}
\usepackage{soul}

\newcommand{\Delete} [1]{\bgroup\noindent\textcolor{red}{\xout{#1}}\egroup\ignorespacesafterend}
\newcommand{\Insert} [1]{\bgroup\noindent\textcolor{blue}{#1}\egroup\ignorespacesafterend}

\newcommand{\twenty}{$20^3\,\upmu$m$^3~$}
\newcommand{\fifty}{$50^3\,\upmu$m$^3~$}
\newcommand{\seventy}{$75^3\,\upmu$m$^3~$}

\begin{document}

\begin{frontmatter}

\title{Modeling the evolution of representative dislocation structures under high thermo-mechanical conditions during Additive Manufacturing of Alloys}

\author[RWTH,JHU]{Markus Sudmanns\corref{cor1}}
\author[NRL]{Athanasios P. Iliopoulos}
\author[NRL]{Andrew J. Birnbaum}
\author[NRL]{John G. Michopoulos}
\author[JHU]{Jaafar A. El-Awady\corref{cor2}}

\cortext[cor1]{Corresponding author: msudmanns@jhu.edu}
\cortext[cor2]{Corresponding author: jelawady@jhu.edu}

\address[JHU]{Department of Mechanical Engineering, The Whiting School of Engineering, The Johns Hopkins University, Baltimore, MD 21218, USA}
\address[NRL]{United States Naval Research Laboratory, Washington DC, 20375, USA}
\address[RWTH]{Digital Additive Production, RWTH Aachen University, Aachen, 52074, Germany}

\begin{abstract}

Mesoscale simulations of discrete defects in metals provide an ideal framework to investigate the micro-scale mechanisms governing the plastic deformation under high thermal and mechanical loading conditions.
To bridge size and time-scale while limiting computational effort, typically the concept of representative volume elements (RVEs) is employed.
This approach considers the microstructure evolution in a volume that is representative of the overall material behavior.
However, in settings with complex thermal and mechanical loading histories careful consideration of the impact of modeling constraints in terms of time scale and simulation domain on predicted results is required.
We address the representation of heterogeneous dislocation structure formation in simulation volumes using the example of residual stress formation during cool-down of laser powder-bed fusion (LPBF) of AISI 316L stainless steel. 
This is achieved by a series of large-scale three-dimensional discrete dislocation dynamics (DDD) simulations assisted by thermo-mechanical finite element modeling of the LPBF process.
Our results show that insufficient size of periodic simulation domains can result in dislocation patterns that reflect the boundaries of the primary cell.
More pronounced dislocation interaction observed for larger domains highlight the significance of simulation domain constraints for predicting mechanical properties.
We formulate criteria that characterize representative volume elements by capturing the conformity of the dislocation structure to the bulk material.
This work provides a basis for future investigations of heterogeneous microstructure formation in mesoscale simulations of bulk material behavior. 

\end{abstract}


\end{frontmatter}


\section{Introduction}
\label{1_Intro}

Multi-scale material modeling involves a paradigm that describes the material behavior on a specific length scale based on an understanding of the physical behavior provided by lower scale models \cite{fish2021mesoscopic}. 
An intelligent homogenization of physical phenomena at lower scales is an important prerequisite for physics-based design of tailored materials in environments that are characterized by physical phenomena whose effects span multiple length scales. 
A particular example is laser-based additive manufacturing of alloys, which requires dealing with highly transient, multi-scale, and multi-physics processes.
While all-atom molecular dynamic (MD) simulations enable modeling different deformation mechanisms in alloys with some relative accuracy, such simulations are severely constrained by the attainable time and length scale and are strongly dependent on the accuracy of the interatomic potential used.
On the other hand, continuum-based approaches, such as crystal plasticity finite element methods (CP-FEM) or continuum theories of dislocations, enable modeling plasticity in relatively large volumes and for relatively larger time-scales. 
However, these simulations are predominately based on phenomenological rules or require inputs about the underlying physical mechanisms, i.e., by coarse graining results from lower length scale models.

Three-dimensional (3D) discrete dislocation dynamics (DDD) simulations are commonly used to investigate the underlying mechanisms governing the evolution of ensembles of crystalline defects and their effect on plasticity and the mechanical properties of crystalline materials \cite{El-Awady2016,Boioli2022}. Such 3D DDD simulations therefore provide a mesoscale modeling framework that bridges the gap in fundamental understanding of the evolution of dislocation plasticity between atomic and continuum computational methods.
In 3D DDD simulations, the number of degrees of freedom increases cubically with the simulation cell edge-length for the same dislocation density. Thus, due to the large number of degrees of freedom involved in 3D DDD simulations, simulations with high dislocation densities for extended plastic strains and relatively large simulation volume are computationally demanding even with advanced parallel computing \cite{raoLargescaleDislocationDynamics2019a}. 
Thus, the concept of ``Representative Volume Elements'' (RVE) is typically employed in 3D DDD simulations where the domain edge-length is commonly limited to $\sim1$ to 20 $\upmu$m and the time scale is limited to hundreds of nanoseconds to microseconds at most, i.e., a limit of $\sim$1-2\% plastic strain. 
For this RVE concept, the simulation cell is defined as a ``primary'' cell of an infinite periodic array of supercells that are exact replicas of this primary cell, i.e., by employing  periodic boundary conditions (PBCs) to the simulation domain \cite{pachaury2022discrete,sudmannsInterplayLocalChemistry2022,voisinNewInsightsCellular2021,bulatov2000periodic,cai2003periodic}.
Examples of utilizing this concept include: simulations of the strain hardening in metals and metal-matrix composites \cite{schulzDislocationdensityBasedDescription2017,grohDiscretecontinuumModelingMetal2004,arsenlisEnablingStrainHardening2007}; quantifying the interactions of dislocations with alloying elements in complex alloys \cite{voisinNewInsightsCellular2021,sudmannsEffectLocalChemical2021,sudmannsInterplayLocalChemistry2022}; or studying the dislocation-precipitate interaction \cite{lehtinenMultiscaleModelingDislocationprecipitate2016a,CHANG2018}; as well as many other applications. The concept is also used for both single crystal simulations \cite{pachaury2022discrete} and polycrystalline simulations \cite{amirmaleki20163d,bagri2018microstructure,BARGMANN2018322}. 
However, it should be noted that the chosen RVE size could inadvertently influence the predictions in the characteristics of the evolving dislocation structure, e.g., spacing and periodicity of dislocation patterns and the subsequent mechanical properties, e.g., yield strength, hardening rate, etc. 
This is commonly attributed to spurious dislocation annihilation/interactions when dislocations interact with their own periodic replicas \cite{devincre2011modeling,madec2004use}.
Thus, care must be taken when interpreting microstructural features/patterns that form with a spatial scale on the order of the chosen RVE simulation domain size, such as in simulations of persistent slip band formation \cite{EL-AwadyPSB2007,ErelPSB2017}, or simulations of laser powder-bed fusion (LPBF) processing of metals, where specific  cellular dislocation structures are observed \cite{bertschOriginDislocationStructures2020,wangAdditivelyManufacturedHierarchical2018}. 

The identification of the appropriate RVE and its characterization in computational modeling has been an active research topic for decades, e.g., for the case of microstructure generation from real experimental data on the level of polycrystals \cite{amirmaleki20163d,bagri2018microstructure,groeber2008framework}, or for composite materials 
\cite{harper2012representativeA,harper2012representativeB}. 
However, on the level of mesoscopic dislocation-based plasticity simulations, these questions have received little attention. 
Existing studies mainly deal with the formulation and implementation of periodic boundary conditions to enforce the continuity of dislocation lines and dislocation flux throughout the deformation history \cite{pachaury2022discrete,bulatov2000periodic} or address the self-annihilation of dislocation lines \cite{devincre2011modeling,madec2004use}. 
On the other hand, attempts to find criteria that would allow for the characterization of required simulation volumes in mesoscale dislocation-based simulations under complex loading scenarios, such as in laser-based additive manufacturing, impose additional challenges associated with the multi-scale and multi-physics nature of the problem.

The complex interplay between the thermal and mechanical environments during the cool-down stage after single track LPBF processing in AISI 316L stainless steels (SS)  \cite{sudmannsInterplayLocalChemistry2022,birnbaumIntrinsicStrainAging2019} includes strong thermal gradients, fast cooling, complex 3D residual stress state, and significant dislocation density multiplication. This leads to significant computational and numerical challenges when modeling the process with 3D DDD simulations. 
Accordingly, this problem provides an ideal setting for quantifying the effects of the chosen 3D DDD simulation cell size on the evolving dislocation structure as well as determining proper criteria that identify the appropriate size.
Here, in order to predict the dislocation structure formation with reasonable accuracy after LPBF of 316L SS, multiphysics FEM simulations are first used to predict the temperature and 3D residual stress profiles using experimental build conditions. 
These results serve as inputs to the subsequent 3D DDD simulations. 
The 3D DDD simulations then focus on quantifying the relationship between the simulation volume sizes, periodic boundary conditions, and the corresponding dislocation structure formation to formulate criteria what would enable the identification of the proper RVE size for the chosen loading scenario.
The 3D DDD simulations employ extraordinarily large cubical simulation cells ranging in size from 20$^3$ to 75$^3$ $\upmu$m$^3$.
In the current context, an RVE refers to a simulation volume that is the smallest structural element of a larger bulk system that still incorporates all relevant characteristics of the overall evolving dislocation structure observed in the bulk system. 
In practical terms, this means a simulation volume that is invariant with respect to the imposed PBCs.
This definition is in alignment with the terminology used in the literature \cite{pachaury2022discrete}. 
The results of these simulations provide an understanding of the requirements for simulation domains in mesoscale simulations as well as strategies for modifying simulation boundary conditions to over come numerical and computational challenges when the needed RVE size is large. 

\section{Methods}
\label{2_Meth}

\subsection{Finite Element model and simulation setup}
\label{sec:FEM}

In order to estimate the stress and temperature profile during cool-down of single-track LPBF 316L SS, FEM simulations were conducted using COMSOL Multiphysics 5.6 \cite{comsol56}.
These profiles represent the input variables into subsequent 3D DDD simulations of the evolution of the dislocation structure.
The concepts underlying the FEM modeling are similar to those described in our earlier study \cite{sudmannsInterplayLocalChemistry2022}, with modifications for improved accuracy, as summarized below.
An isotropic thermo-mechanical coupled model is applied to an idealized single-track LPBF process of 316L SS neglecting the influence of the powder-bed and details about variations in the chemical composition.
The laser input was modeled as a surface heat flux moving in the scanning direction, where the intensity was chosen to provide a melt pool size and shape close to that observed experimentally \cite{birnbaumIntrinsicStrainAging2019}. 
The FEM model employs a routine accounting for the thermo-elasto-plastic conditions during the phase transformation induced by the melting then re-solidification and the resulting induced thermal strains.
Because of the need to solve the structural problem, the model is based on a Lagrangian approach, which couples structural mechanics and heat transfer physics. 
A custom developed temperature-dependent plasticity model was used, which incorporates the deactivation of plastic strain accumulation within the melt-pool volume when the temperature exceeds the melting temperature of 1700K. The accumulation of plastic strain is reactivated upon solidification. 
Furthermore, the increase in the thermal strains during cool-down was accounted for by changing the reference temperature from room temperature to the melting temperature of 1700K inside the resolidified material upon solidification.
Details of the model are described in Supplementary Section S1.

The mechanical properties selected for these simulations are those for 316L stainless steel, which were compiled from multiple literature sources \cite{Lee1998, CHEN2006229, KARKALOS2018107,Ledbetter1981} and heat transfer properties from \cite{osti_4152287}. For a detailed description of the mechanical and thermal properties we refer to \cite{sudmannsInterplayLocalChemistry2022}.
The yield stress was assumed to be temperature and strain rate dependent and was modeled using a Johnson-Cook model with material constants identified in \cite{KARKALOS2018107}. 

The mechanical boundary conditions of the FEM simulation domain, shown in Supplementary Fig.\,S1, were fixed at the bottom surface at $Z = 0$, while symmetry boundary conditions were employed on the side surface at $Y=0$ using the $XZ$-plane as a mirror plane. For the symmetry boundary, $-\mathbf{n} \cdot \mathbf{q} = 0$ was enforced for the thermal problem and $\mathbf{v} \cdot \mathbf{n}=0$ was enforced for the mechanical problem. Here, $\mathbf{q}$ is the heat flux vector, $\mathbf{v}$ is the displacement vector, and $\mathbf{n}$ is the normal vector to the surface. A convective heat flux boundary condition was applied using $-\mathbf{n} \cdot \mathbf{q} = h (T_{\infty} - T)$ to simulate the presence of material at all except the top and the symmetry boundaries, where $T_{\infty}$ is the temperature of the environment, i.e. the room temperature, and $h$ is the heat transfer coefficient. An approximate value for $h$ was identified as $h=500\,$W$/($m$^2$K$)$ \cite{Iliopoulos2015-ro}.  

The heat input by the laser beam was accounted for in the form of a heat flux boundary condition applied to the top boundary, which moves along the negative $X$-direction with a laser beam velocity of $v_\mathrm{laser}=0.9\,$m/s.
The heat flux is given by
\begin{equation}
	-\mathbf{n} \cdot \mathbf{q} = a P_0 f\left( \mathbf{O}, \mathbf{e} \right) \frac{\left| \mathbf{e} \cdot \mathbf{n}\right|}{\lVert \mathbf{e} \rVert},
\end{equation}
where $a = 0.65$ is the laser coupling coefficient, $P_0 = 370\,$W is the laser power as in \cite{birnbaumIntrinsicStrainAging2019}, and $\mathbf{e}=\left\{ 0\mathbf{i},0\mathbf{j}, -1 \mathbf{k}\right\}^T$ is the beam orientation vector. The function $f$ defines the beam shape and was assumed to be of a Gaussian form according to:
\begin{equation}
	f\left(\mathbf{O},\mathbf{e}\right) = \frac{1}{2 \pi \sigma^2} e^{-\frac{s^2}{2 \sigma^2}},
\end{equation}
with
\begin{equation}
	s=\frac{\lVert \mathbf{e} \times \left(  \mathbf{x} - \mathbf{O}\right)\rVert}{\lVert \mathbf{e} \rVert},
\end{equation}
where $\mathbf{x}$ denotes the coordinates of the points along the top surface at $y = 0$ and $\mathbf{O}$ is the center of the beam application. 
The Gaussian distribution $\sigma$ was assumed to be equal to $d/4$ where $d$ is the apparent beam diameter equal to $\SI{132}{\micro \meter}$.

\subsection{Discrete Dislocation Dynamics simulation setup}
\label{sec:DDD_setup}

Large scale 3D DDD simulations allow for a detailed study of the microstructure evolution, which are driven by thermo-mechanical residual stresses induced by the cool-down of LPBF processing on dislocation level.
All DDD simulations conducted in this work employ an in-house modified version of the 3D DDD open-source code ParaDiS \cite{arsenlisEnablingStrainHardening2007}, which is modified to allow for atomistically informed cross-slip mechanisms \cite{husseinMicrostructurallyBasedCrossslip2015}.

To adequately model the dislocation microstructure evolution in 316L SS alloys, a physically meaningful representation of the dislocation mobility and interaction in the chemical environment of 316L involving solute microsegregation was considered.
The correct dislocation mobility including atomic misfit stresses and their spatial variations induced by the alloying elements are derived from the elastic interaction between the solutes and a dislocation segment according to \cite{varvenneTheoryStrengtheningFcc2016,varvenneSoluteStrengtheningRandom2017}. 
In this model, the critical resolved shear stress (CRSS) for a dislocation segment is derived from the elastic interaction between the solutes and a dislocation segment. It can be quantified by an energy barrier $\Delta E_b$, which the dislocation must overcome after reaching the zero-temperature strength $\tau_c^0$.
By accounting for thermal activation, a temperature-dependent CRSS of the dislocation can be described as in \cite{leysonSoluteStrengtheningHigh2016}
A detailed explanation of the model and its implementation in 3D DDD can be found in \cite{sudmannsEffectLocalChemical2021,sudmannsInterplayLocalChemistry2022}.
The shear modulus and the Poisson ratio used in the DDD model are temperature-dependent and have been compiled from literature \cite{ChaudhuryAtlas,desu2016mechanical,Byun2017a}.
Furthermore, the dislocation drag coefficient in the dislocation mobility law was chosen according to an estimation for stainless steels which has recently been obtained from molecular dynamics (MD) simulations for Fe$_{0.7}$Ni$_{x}$Cr$_{0.3 - x}$ \cite{chuTemperatureCompositionDependent2020} using $x=0.15$.

We conducted a set of large-scale 3D DDD simulations mimicking the temperature-dependent microstructure evolution during cool-down of LPBF 316L SS.
Since the DDD simulations focus only on the solid state of the material and 316L SS solidifies when the melt pool temperature drops below $T_\mathrm{melt} = 1700$K, all the current DDD simulations take only as input the temperature and stress profiles predicted from the FEM simulations at the instance the temperature drops below $T_\mathrm{melt}$, which is designated as time $t = 0$ in the DDD simulations. 
All simulations capture dislocation plasticity during the cool-down from 1700K to 1400K.
Although these assumptions are a simplification of the solidification processes in real alloys, they are considered adequate for the purpose of this study, which aims at a qualitative representation of the plasticity at early LPBF cooldown in small sections of the material. 
Thus, a relatively low initial dislocation density is assumed to mimic the expected defect densities at very high temperatures right after solidification \cite{Blackwell1967,INOUE1974,NAKANO2011,YI2019}.

A statistical analysis on randomly generated microstructures in Supplementary Section S2 shows a clear relationship between dislocation density and system volume size for generating microstructures that are statistically representative in very low dislocation density regimes.
Therefore, three extraordinarily large cubic simulation cell sizes of $20^3\,\upmu$m$^3$, $50^3\,\upmu$m$^3$, and $75^3\,\upmu$m$^3$, are chosen for the subsequent analysis to generate physically reasonable initial dislocation line lengths.
We populated the system with randomly positioned and oriented dipolar dislocation loops with edge lengths of $1\,\upmu$m and dipole heights within a range of 250 nm and 500 nm, to obtain an initial dislocation density of $1.2\times10^9\,$m$^{-2}$, $9.9\times10^8\,$m$^{-2}$, and $6.5\times10^8\,$m$^{-2}$, in the $20^3\,\upmu$m$^3$, $50^3\,\upmu$m$^3$, and \seventy simulation cell sizes, respectively.
To allow for statistically comparable conditions we thus ensured that in the \twenty simulation cell size all slip systems are populated initially.

\begin{figure}[h]
    \centering
    \includegraphics[width=0.5\textwidth]{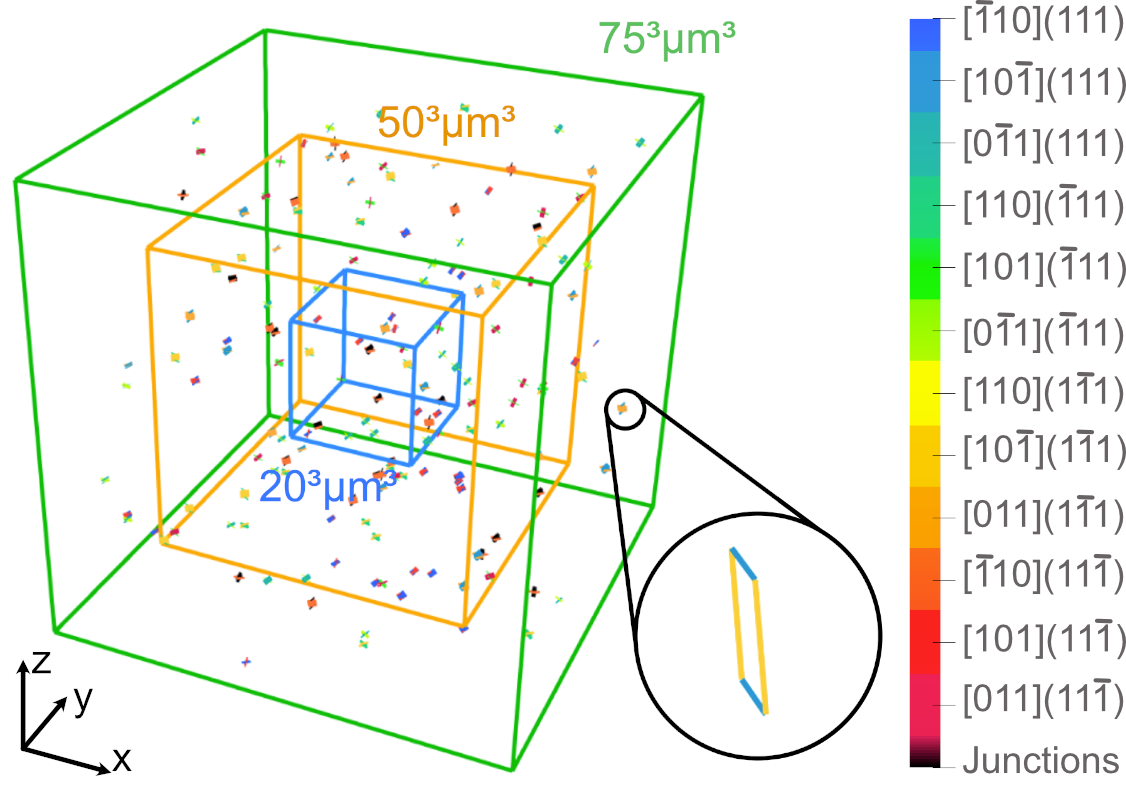}
    \caption{Representation of the cubic simulation cells used in the current study showing the initial randomly distributed and oriented dipolar dislocation loops used in the different simulation cells. The colored boxes represent the three PBC simulation cells used in this study. The colored lines segments represent the dislocations and their colors represent the slip system they belong to.}
    \label{fig:Initial_density}
\end{figure}

All DDD simulations incorporate periodic boundary conditions for the dislocation motion to mimic a single-crystalline bulk material.
A visualization of the $20^3\,\upmu$m$^3$, $50^3\,\upmu$m$^3$, and \seventy simulation cells superimposed by the initial distribution of dipolar dislocation loops is shown in Fig.\,\ref{fig:Initial_density}.
The \seventy simulation cell is defined as the reference case for comparison with the \twenty and the \fifty simulation cell.
In all DDD simulations, the $X$, $Y$, and $Z$ axes correspond to the $[100]$, $[010]$, and $[001]$ crystal axes, respectively.

All simulations further contained a cellular segregation to match that observed experimentally \cite{sudmannsInterplayLocalChemistry2022}. This is achieved by assuming a higher concentration of Mo, Cr, and Mn in the chemical cell walls as compared to the cell interior, in agree ment with experimental observations.
The chemical cell walls are oriented parallel to the [100] and [010] plane with a 800\,nm spacing according to experimental observations \cite{wangAdditivelyManufacturedHierarchical2018,liuDislocationNetworkAdditive2018}. 
For details on the modeling of solute segregation within the DDD framework including the chemical composition we refer the reader to our earlier work \cite{sudmannsInterplayLocalChemistry2022}. 

To identify periodic dislocation patterns in the dislocation structure predicted by the DDD simulations using $20^3\,\upmu$m$^3$, $50^3\,\upmu$m$^3$, and $75^3\,\upmu$m$^3$ simulation cells, we use three-dimensional correlation analyses of the voxel-averaged dislocation density. Further, this method is used to quantify similarities in the predicted dislocation structures. A detailed explanation of the process is given in \ref{sec:Correlation}.

\subsection{Computing the dislocation flux}
\label{sec:Method_flux}

In DDD simulations the predicted microstructure evolution and resulting material properties originate from effects related to dislocation motion.
Thus, we postulate that a methodology to characterize the ability of a certain simulation volume in DDD to capture the actual behavior of the system can be formulated by evaluating the flux of dislocations across boundaries of computational \textit{sub-volumes} embedded in a bulk microstructure.
Here, the term ``sub-volume'' refers to a section of the dislocation structure that was extracted from the reference cell.
We define the total flux of dislocations \textit{out of} a sub-volume, $f_\mathrm{out}$, as the line length $\Delta l_\mathrm{out}$ leaving a volume through a surface area $A$ within a time increment $\Delta t$:
\begin{equation}
f_\mathrm{out} = \frac{\Delta l_\mathrm{out}}{A\Delta t},
\label{eq:flux_out}
\end{equation}
and correspondingly for the total flux of dislocations \textit{into} a sub-volume is
\begin{equation}
f_\mathrm{in} = \frac{\Delta l_\mathrm{in}}{A\Delta t},
\label{eq:flux_in}
\end{equation}
 where $\Delta l_\mathrm{in}$ is the dislocation line length leaving a volume through a surface area $A$. Here, the area $A$ corresponds to the total surface area of the cubic sub-volume of interest.

An ideal size-invariant RVE of a bulk material is characterized by a microstructure evolution that is qualitatively independent of the presence of computational boundaries or the size of the chosen sub-volume.
Based on the evaluation of dislocation fluxes as defined above, this condition is fulfilled when the dislocation flux leaving the sub-volume is equal to the dislocation flux entering the sub-volume.
A further increase in sub-volume size should then not lead to a qualitative difference in microstructure formation.
Using these measures, we define an absolute flux difference as 
\begin{equation}
\Delta f = |f_\mathrm{out} - f_\mathrm{in}|
\label{eq:Delta_flux}
\end{equation}
and a relative flux difference as
\begin{equation}
\Delta \bar{f} =  \frac{|f_\mathrm{out} - f_\mathrm{in}|}{\mathrm{max}\langle f_\mathrm{out},f_\mathrm{in}\rangle} = \frac{\Delta f}{\mathrm{max}\langle f_\mathrm{out},f_\mathrm{in}\rangle},
\label{eq:Delta_rel}
\end{equation}
where we normalize the absolute flux difference by the maximum $f_\mathrm{out}$ or $f_\mathrm{in}$ in the given time increment $\Delta t$ in order to obtain a value within the interval $[0,1]$.
Consequently, we define size-invariant periodicity as net-zero relative dislocation flux, i.e. $\Delta \bar{f} = |f_\mathrm{out} - f_\mathrm{in}|/\mathrm{max}\langle f_\mathrm{out},f_\mathrm{in}\rangle = 0$.
This implies that either no flux difference occurs, i.e. $\Delta f = 0$, or is small compared to the total flux, i.e. $\mathrm{max}\langle f_\mathrm{out},f_\mathrm{in}\rangle >> \Delta f$.
The opposite case, i.e. $\Delta \bar{f} = 1$, describes a condition in which the net dislocation flux equals the total dislocation flux in any direction, which means that either dislocations only enter the sub-volume or only leave the sub-volume in the given time increment.
This implies that a strong heterogeneity in bulk microstructure evolution must exist, which the chosen sub-volume is unable to represent. 
The relative flux difference defined by Eq.\,(\ref{eq:Delta_rel}) can therefore be understood physically as a quantity that relates the net dislocation flux to the overall dislocation activity, and thus, evaluates the capability of the chosen sub-volume to capture the essence of microstructure formation in the bulk material.
We note that this definition intentionally neglects the dislocation line length increase due to multiplication within the sub-volume in order to be compatible with the definition of periodic boundary conditions in typical DDD simulation frameworks.

\begin{figure}[h]
    \centering
    \includegraphics[width=\textwidth]{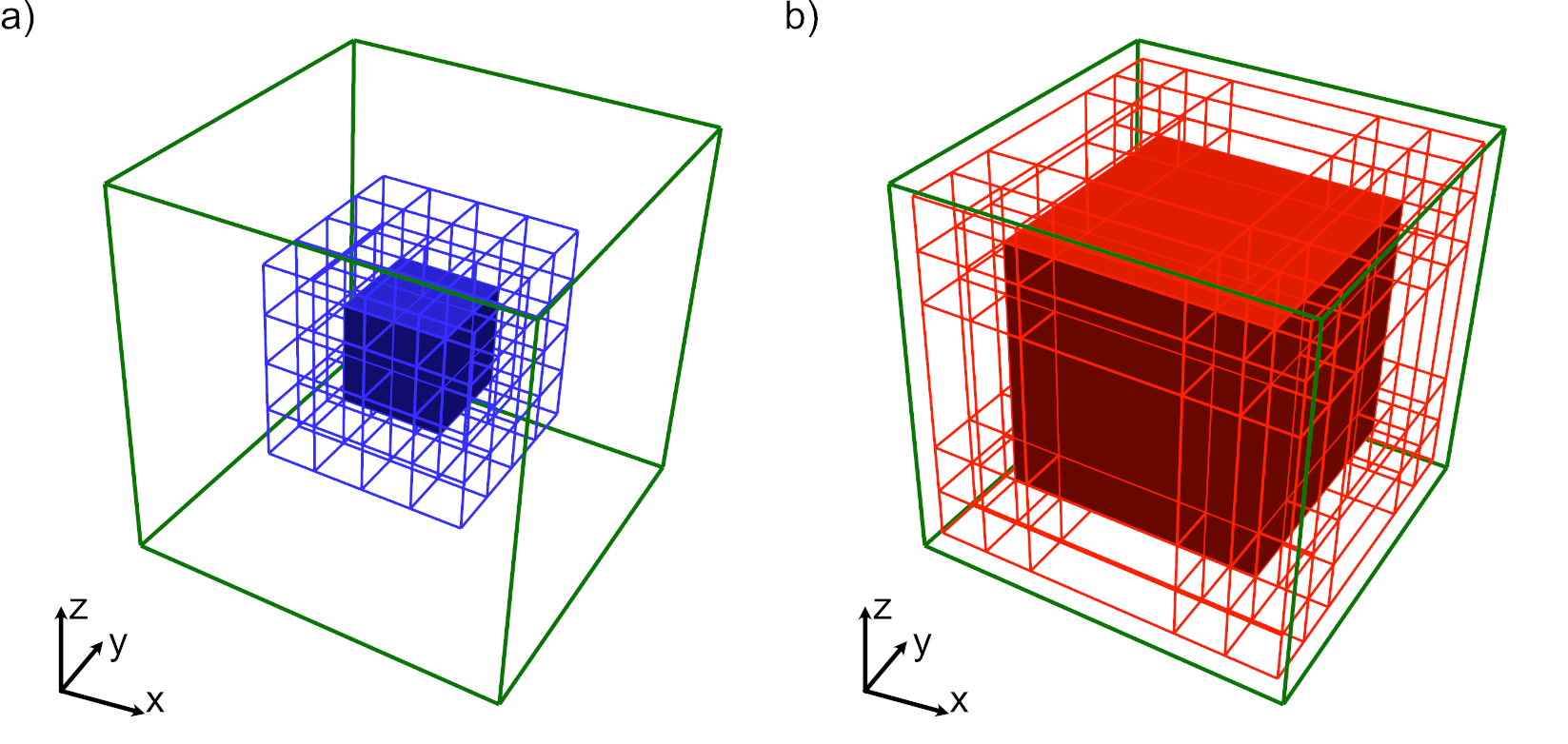}
    \caption{Schematics of the cubic sub-volumes, used for evaluation of the dislocation flux, having size: (a) $20^3\,\upmu$m$^3$; and (b) $50^3\,\upmu$m$^3$, inside the \seventy volume, which is shown as a green wire-frame in (a) and (b). The different \twenty and \fifty sub-volumes translated by $\Delta X, \Delta Y, \Delta Z = \pm 10\,\upmu$m in each directions are shown as blue and red wire-frames in (a) and (b), respectively.}
    \label{fig:Flux_methods}
\end{figure}

We applied the method to the DDD simulation using the \seventy simulation cell as explained in Section \ref{sec:DDD_setup}, which will act as “bulk-material” for the purpose of the study.
We evaluated the dislocation flux across boundaries of imaginary sub-volumes with \twenty and \fifty sizes inside the \seventy DDD simulation volume to identify the relationship between the sizes of sub-volumes and the RVE condition as defined in this section.
To investigate the heterogeneity in microstructure formation across the volume, the dislocation flux was evaluated at different positions across the volume by translating the cubic \twenty and \fifty sub-volumes by $\Delta X, \Delta Y, \Delta Z = \pm 10\,\upmu$m from its original position at the center of the \seventy volume.
This is shown schematically in Fig.\,\ref{fig:Flux_methods}, where the positions of the translated sub-volumes are indicated by blue and red wireframes.
These investigations aim at demonstrating the presented method as a criteria for representativity of chosen sub-volume sizes acting as a RVE in DDD simulations. 

%

\section{Results}
\label{3_Res}

\subsection{The temperature and 3D residual stress profiles during cool-down}

\begin{figure}[h]
    \centering
    \includegraphics[width=\textwidth]{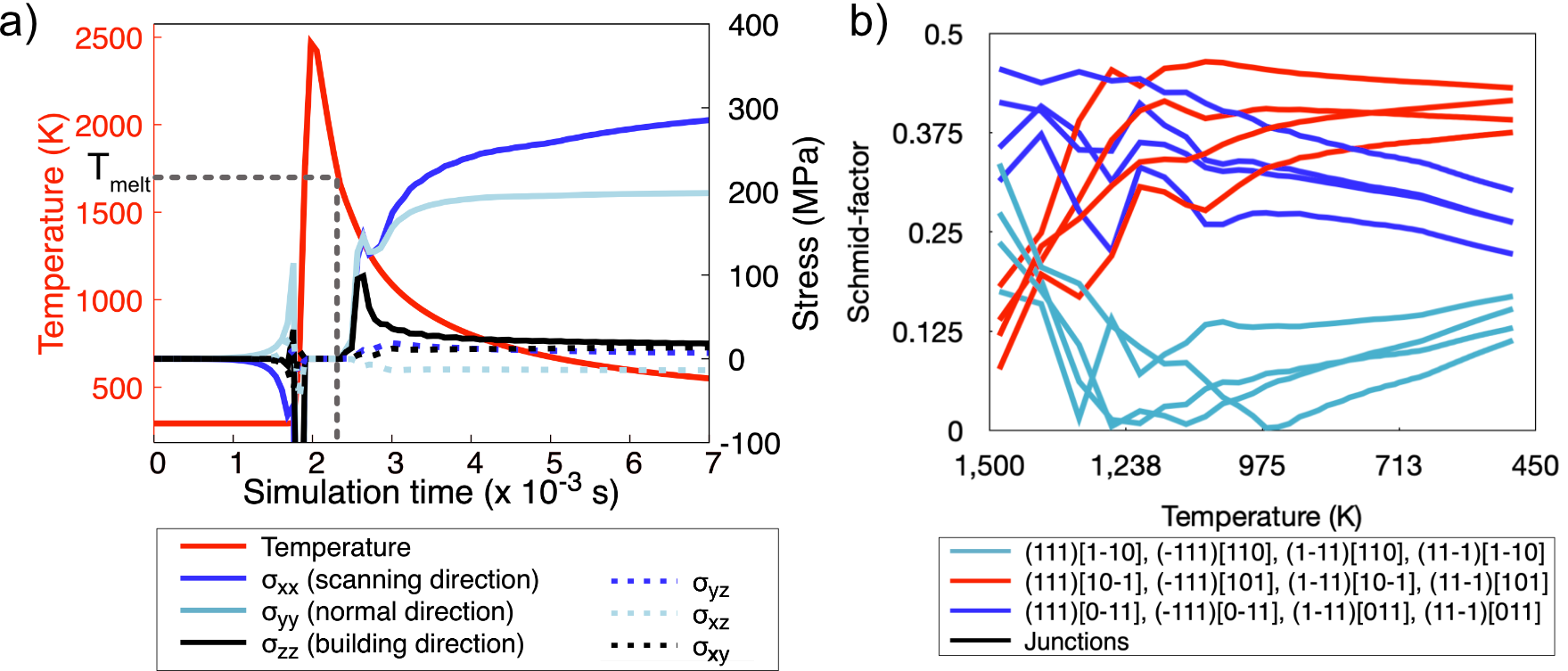}
    \caption{(a) The predicted time evolution of the temperature and stress components from FEM simulations of single-track LPBF scan of 316L SS for a point inside the former melt-pool (see supplementary Fig.\,S1) and (b) the resulting Schmid-factors on all slip systems (bundled into groups of four according to their overall behavior; the split-up into all slip systems is shown in supplementary Fig.\,S3) as a function of temperature at the same point in (a). The melting temperature of 316L SS at 1700K is indicated in (a) for reference.}
    \label{fig:1_FEM}
\end{figure}

The first challenge to overcome is to evolute the temperature and residual stresses that develop during the LPBF process. The temperature and the components of the residual stress tensor as a function of time predicted for a point inside the former melt-pool (see supplementary Fig.\,S1) after solidification and during cool-down are shown in Fig.\,\ref{fig:1_FEM}(a). These results are predicted from thermo-mechanical coupled FEM simulations of a one way single-track LPBF scan of 316L SS incorporating a formation of the thermal stresses induced by the re-solidification and cool-down of the material as described in section\,\ref{sec:FEM}. It is observed that the temperature rapidly decreases after solidification (melting temperature for 316L SS is 1700K) and this is accompanied by a significant increase in the normal stresses up to 300\,MPa and 200\,MPa parallel, $\sigma_\mathrm{xx}$, and transversal, $\sigma_\mathrm{yy}$, to the scanning direction, respectively. 
Notably, an initial peak stress can be observed in all three normal stress components in the early stages of cool-down at $\sim1350K$.
This peak stress can be attributed to the formation of a stress wave emanating from the melt-pool boundary.
However, the normal stress in the build direction, $\sigma_\mathrm{zz}$, then rapidly decreases to zero below $1300$K.
It should be noted that these residual stresses originate from contraction constraints during the cool-down stage, which is enforced by the unmelted material. This agrees well with previous FEM simulations on single-track LPBF AISI 316L SS \cite{sudmannsInterplayLocalChemistry2022} and experimental measurements on bulk LPBF AISI 316L SS samples \cite{simson2017residual}.

%

To provide a first estimate of the relation between the temperature, the thermo-mechanical stress profiles, and the expected plastic slip activity after solidification, the nominal Schmid factors on each of the 12 FCC slip systems were calculated using the MTEX Matlab package \cite{bachmann2010texture}, and shown in Fig.\,\ref{fig:1_FEM}(b). For simplicity, in these calculations the $[100]$, $[010]$, and $[001]$ crystal directions are assigned  parallel to the $x$ (scanning), $y$ (normal), and $z$ (build) axes, respectively.
The calculated Schmid-factors can be bundled into three groups with four slip systems in each. 
It is observed that four slip systems (colored in light blue in Fig.\,\ref{fig:1_FEM}(b)) start with relatively high Schmid-factors and then drop significantly, whereas 4 other slip systems (colored in dark blue in Fig.\,\ref{fig:1_FEM}(b)) also start with high Schmid-factors, but then only drop moderately. 
The remaining four slip systems (colored in red in Fig.\,\ref{fig:1_FEM}(b)) start with low Schmid-factors and gradually increase to become more dominant at lower temperatures.
Notably, the Schmid-factor of two bundles intersect at around 1100K, indicating a change in nominal slip system activity.

\subsection{The effect of the simulation cell size on the evolution of dislocation microstructures during cool-down}

\begin{figure}[h]
    \centering
    \includegraphics[width=\textwidth]{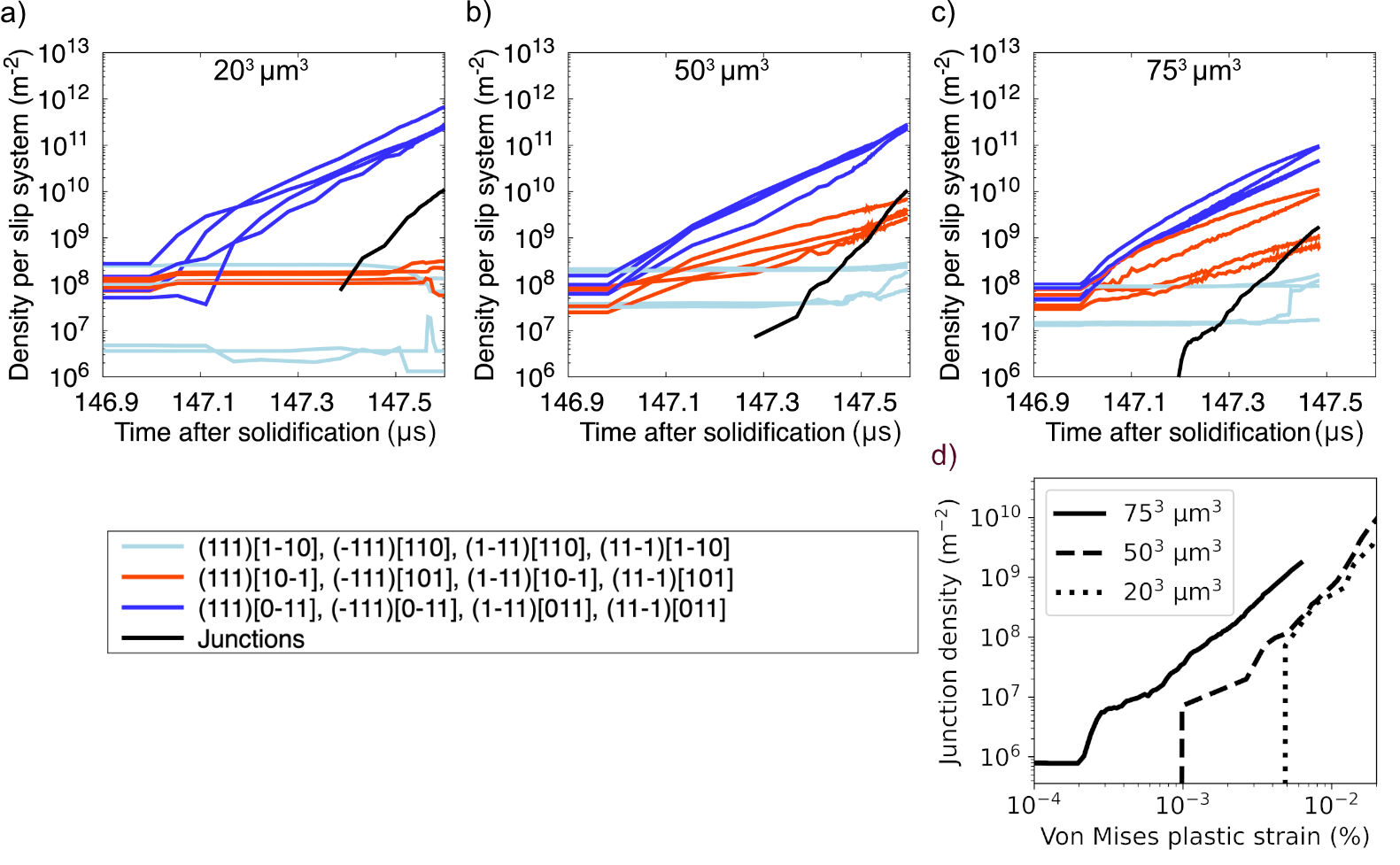}
    \caption{Evolution of the dislocation density versus time on the twelve FCC slip systems (bundled into groups of four according to their overall behavior; the split-up into all slip systems is shown in supplementary Fig.\,S4) as well as that for junctions for the: (a) $20^3\,\upmu$m$^3$; (b) $50^3\,\upmu$m$^3$; and (c) \seventy simulation cells. The simulation time $t = 0$ represents the start of the DDD simulations, which coincide with the instance of solidification. The evolution of the density of dislocation junctions in all three simulations as a function of von Mises plastic strain, $\varepsilon_\mathrm{pl}^\mathrm{VM}$, is shown in (d).}
    \label{fig:3_Density_evolution}
\end{figure}

Figure\,\ref{fig:3_Density_evolution}(a)-(c) shows the predicted time evolution of the dislocation density per slip-system for the three extraordinarily large cubic simulation cell sizes of $20^3\,\upmu$m$^3$, $50^3\,\upmu$m$^3$, and $75^3\,\upmu$m$^3$.
It is observed that the rate of dislocation density increase on slip systems $(111)[10\bar{1}]$, $(\bar{1}11)[101]$, $(1\bar{1}1)[10\bar{1}]$, and $(11\bar{1})[101]$ (colored in red) are drastically different between the \twenty simulation cell and that in the \seventy and \fifty simulation cell. In the \twenty cell case these slip systems are relatively inactive for the temperature range modeled here. This is in contrast to the slip system activity expected from the nominal Schmid-factor calculations shown in Fig.\,\ref{fig:1_FEM}(b) in this temperature range.
On the other hand, the density evolution in the \seventy and \fifty cells is in agreement with the slip system activity expected.

%
%

The differences in the evolution of the dislocation density on the different slip systems for the three simulation cell sizes leads to differences in dislocation interactions and subsequently in junction formation.
This is evident by Fig.\,\ref{fig:3_Density_evolution}(d), which shows the evolution of the density of sessile dislocation junctions (i.e. Hirth and Lomer junctions) as a function of von Mises plastic strain $\varepsilon_\mathrm{pl}^\mathrm{VM}$ for the three simulation cell sizes.
A higher density of junctions in the \seventy sub-volume simulation cell is observed as compared to the \twenty cell for the same level of von Mises plastic strain.
It should be noted that the slip systems showing the highest dislocation density increase in Fig.\,\ref{fig:3_Density_evolution}(a)-(c) represent combination of slip systems that promote collinear and Hirth junction formation. Thus, they only weakly contribute to the formation of a strong dislocation network.
In contrast, the additional slip system bundles that are only active in the \fifty and \seventy simulations cells contribute to glissile and lomer junction formation. 
It can thus be expected that, in contrast to the \fifty and \seventy simulation cells the junction density in the \twenty simulation cell mostly consists of weak Hirth-type junctions.

\begin{figure}[h]
    \centering
    \includegraphics[width=\textwidth]{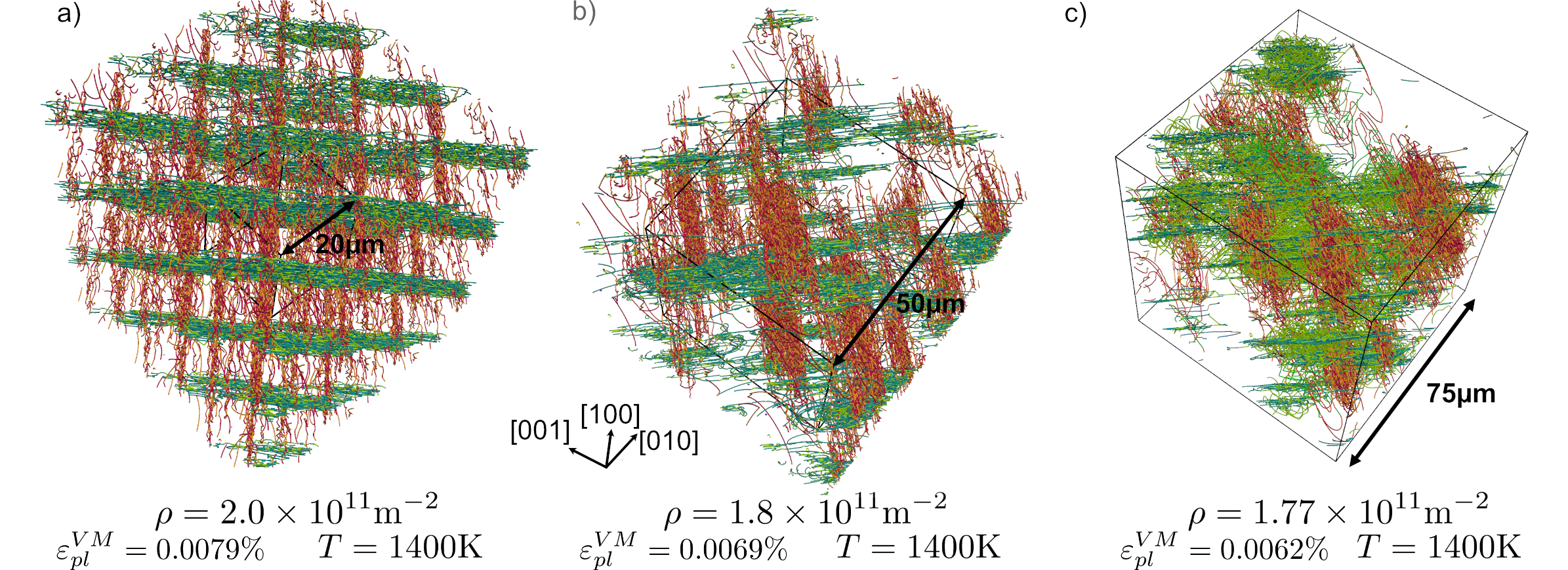}
    \caption{3D view of the dislocation structure as predicted by 3D DDD simulations after cool-down from 1700K to 1400K using a: (a) $20^3\,\upmu$m$^3$; (b) $50^3\,\upmu$m$^3$; and (c) $75^3\,\upmu$m$^3$ simulation cells. For a better comparison, the dislocation structure in the case of the \twenty and \fifty simulation cells are periodically replicated within a \seventy volume. The black lines indicate the edges of the actual periodic simulation cells. The dislocation lines are colored according to the slip systems. The total dislocation density, $\rho$, and the von Mises plastic strain, $\varepsilon_\mathrm{pl}^\mathrm{VM}$, for each case is also shown. The 3D views of the periodic replicas for the three simulation cells are also shown in Supplementary Fig.\,S5.}
    \label{fig:4_Dislocation_microstructure_3D}
\end{figure}

The 3D dislocation structure predicted using the $20^3\,\upmu$m$^3$, $50^3\,\upmu$m$^3$, and \seventy simulations cell after cool-down from 1700K to 1400K are shown in Fig.\,\ref{fig:4_Dislocation_microstructure_3D}.
The dislocation structures in the \twenty and \fifty simulation cells are periodically replicated within a \seventy volume for better comparison between the three dislocation structures.
Clearly, a very distinct periodic pattern-like dislocation structure is evident in the case of the \twenty simulation cell, whereas those predicted in the \fifty and \seventy simulation cells are only loosely arranged in a regular periodic arrangements.

%

\begin{figure}[h]
    \centering
    \includegraphics[width=\textwidth]{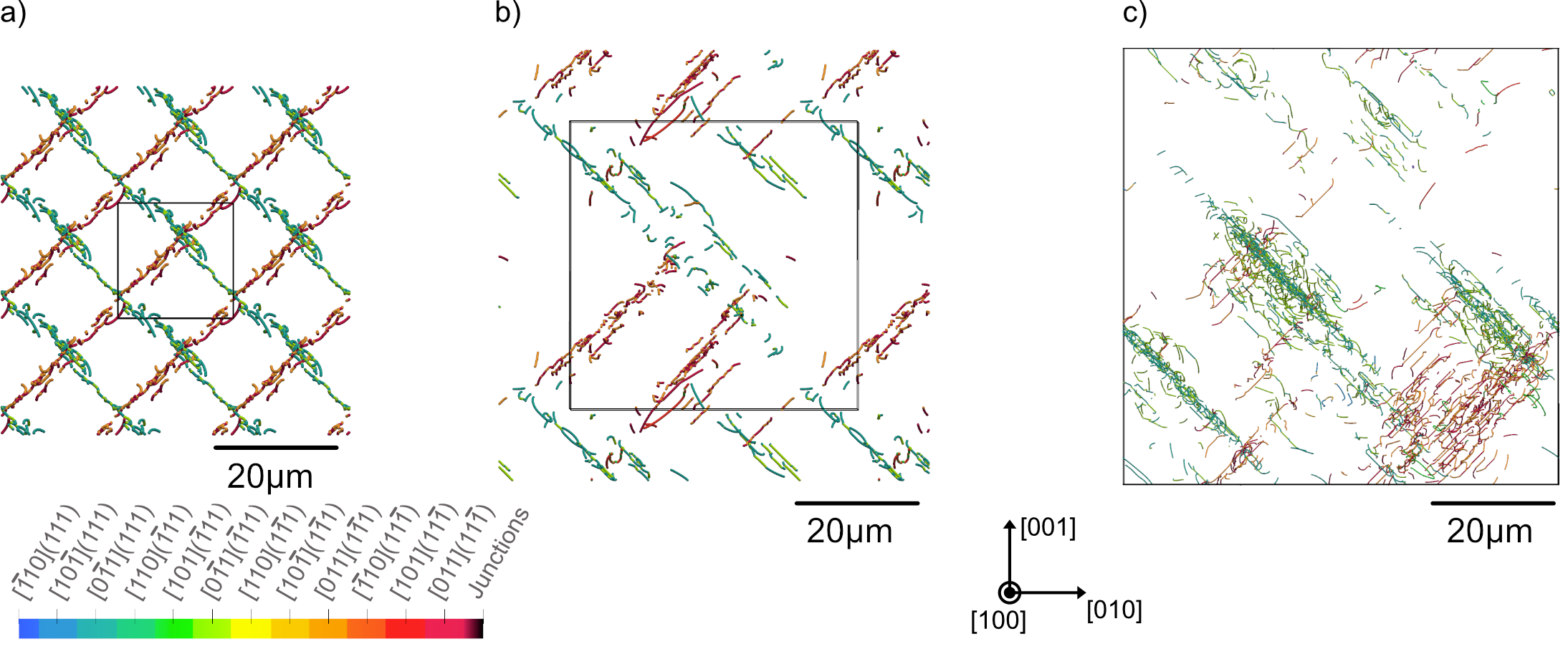}
    \caption{The dislocation structure in TEM-like foils, having a thickness of $1\,\upmu$m perpendicular to the $[100]$-direction, and extracted from the center of the 3D DDD simulation cells shown in Fig.\,\ref{fig:4_Dislocation_microstructure_3D} for the: (a) $20^3\,\upmu$m$^3$; (b) $50^3\,\upmu$m$^3$; and (c) \seventy simulation cells. The periodic replicas of the \twenty and \fifty simulation cells are shown to match the dimensions of the \seventy simulation cell for better comparison. The black lines indicate the boundaries of the actual periodic simulation cell.}
    \label{fig:5_Dislocation_microstructure_periodic}
\end{figure}

For a closer analysis and quantification of the effect of the simulation cell on the predicted dislocation structure, Fig.\,\ref{fig:5_Dislocation_microstructure_periodic} shows transmission electron microscopy (TEM)-like foils, having a thickness of $1\,\upmu$m perpendicular to the $[100]$-direction, which were extracted from the center of each simulation cell.
The \twenty simulation cell shows uniform and periodic dislocation walls, which form very regular dislocation cells when accounting for the periodicity of the simulation cell.
The edge length of the formed dislocation cells correspond to the dislocation wall spacing and are on the order of $\sim{15}\,\upmu$m.
It can be easily inferred that these regular dislocation patterns reflect the imposed periodic boundaries, rather than being naturally forming structures.
In contrast, the dislocation structures in the \fifty and even more evidently in the \seventy simulation cells appear more random with two different dislocation wall spacings of $\sim{5}\,\upmu$m and $\sim{20}\,\upmu$m in the \fifty simulation cell and $\sim{7}\,\upmu$m and $\sim{30}\,\upmu$m for the \seventy simulation cell.
These observations hold also for the corresponding dislocation structure observed in the TEM-like foils extracted near the top and the bottom of each simulation cell, as shown in Supplementary Fig.\,S6 - Fig.\,S8.

\begin{figure}
    \centering
    \includegraphics[width=0.9\textwidth]{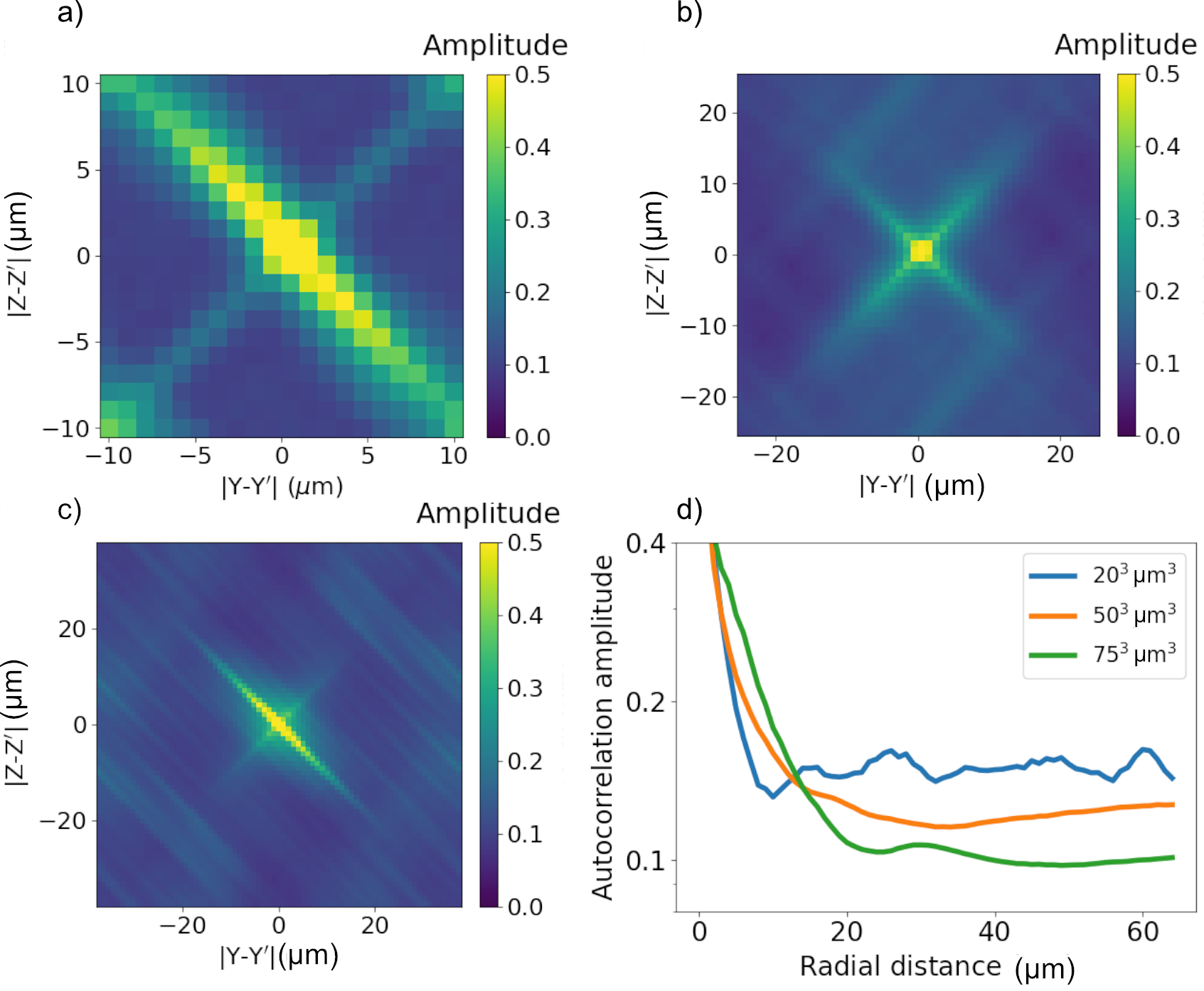}
    \caption{2D contour plots showing the amplitudes of the 3D autocorrelation function from the voxel-averaged dislocation density of the TEM-like foils shown in Fig.\,\ref{fig:5_Dislocation_microstructure_periodic} for the: (a) $20^3\,\upmu$m$^3$; (b) $50^3\,\upmu$m$^3$; and (c) \seventy simulation cells. (d) The radially averaged 3D autocorrelation profile from the center of each simulation-cell.}
    \label{fig:microstructure_autocorrelation}
\end{figure}

To quantify the periodicity in the formed dislocation patterns in each simulation cell, the normalized 3D autocorrelation functions are calculated for the voxelized dislocation density in each simulation cell.
The data for this analysis were prepared by calculating local dislocation densities within voxels of $1\,\upmu$m edge length throughout the entire volume. 
The spatial autocorrelation then quantifies the correlation between this dataset and a translated copy that is mapped on itself.
For a detailed derivation, we refer to \mbox{\ref{sec:Correlation}}.
The amplitude of the normalized 3D autocorrelation functions for the TEM-like foils extracted from the center of the $20^3\,\upmu$m$^3$, $50^3\,\upmu$m$^3$, and \seventy simulation cells, are shown as 2D contour plots in Fig.\,\ref{fig:microstructure_autocorrelation}(a)-(c).
A strong periodicity is evident in the case of the \twenty simulation cell, which reflects its periodic dimensions.
However, the periodicity is less evident in the case of the \fifty and \seventy simulation cells.
This effect can also be seen by radially averaging the 3D autocorrelation functions from the center of the respective cells, as shown in Fig.\,\ref{fig:microstructure_autocorrelation}(d).
Here, the amplitude in the \twenty simulation cell shows various peaks along the radial distance, confirming the periodicity in the predicted dislocation structure in the smallest simulation cell.
In contrast, no relevant periodicity is observed in the \fifty or \seventy simulation cells, indicating a transition towards a more random heterogeneous dislocation structure.

%
%

\begin{figure}
    \centering
    \includegraphics[width=0.9\textwidth]{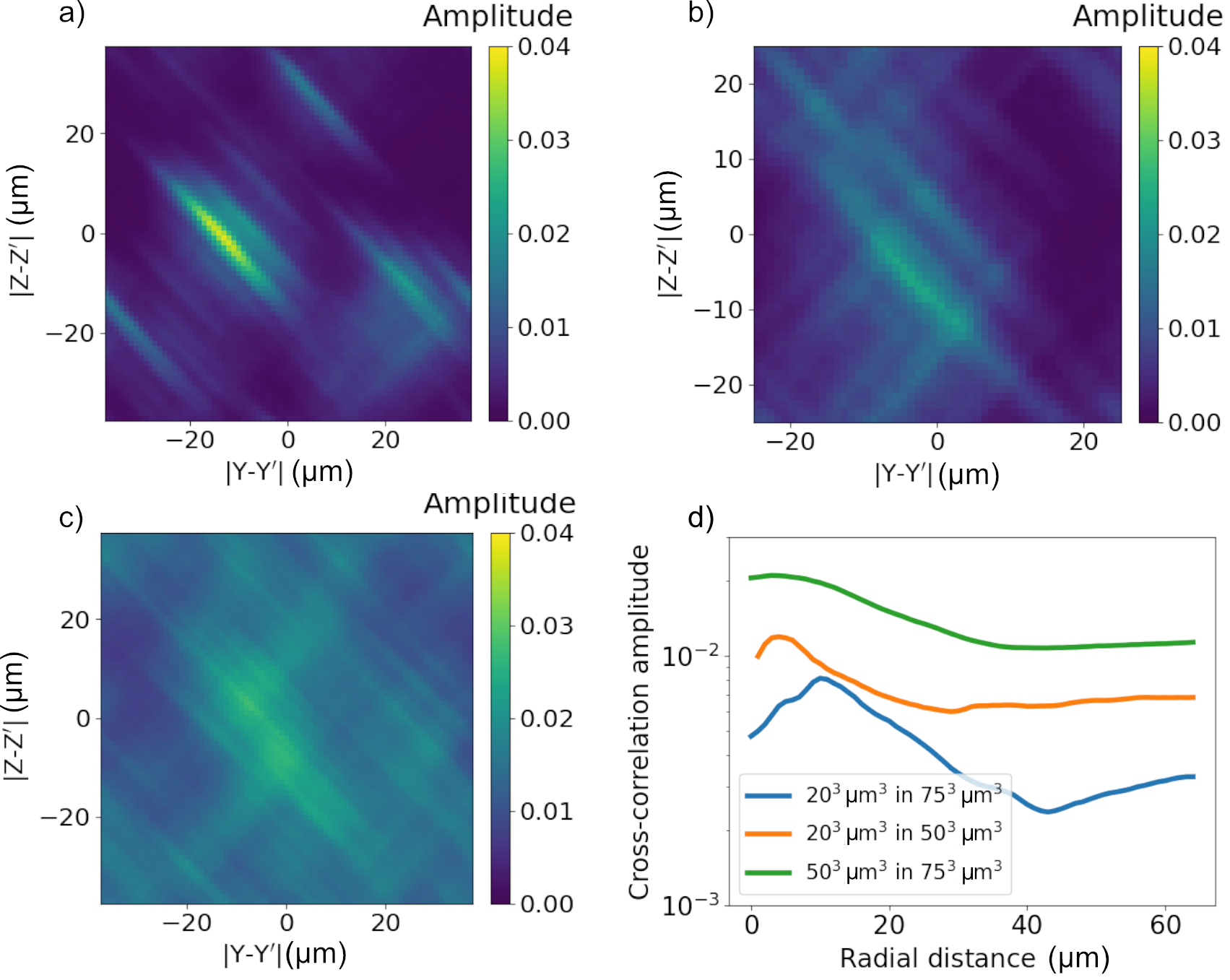}
    \caption{2D contour plots showing the amplitudes of the 3D cross-correlation function from the voxel-averaged dislocation density for the TEM-like foils shown in Fig.\,\ref{fig:5_Dislocation_microstructure_periodic} for the: (a) \twenty - $75^3\,\upmu$m$^3$; (b) \twenty - $50^3\,\upmu$m$^3$; and (c) \fifty - \seventy simulation cell pairings. (d) Shows the radially averaged 3D autocorrelation profile from the center of each simulation cell pairing.}
    \label{fig:microstructure_crosscorrelation}
\end{figure}

The similarity between the dislocation structures predicted in each of the three simulation cells is also characterized by computing the cross-correlation of the voxelized dislocation density in each cell. Here, sections of an image are compared with a given template (see Section \ref{sec:Correlation}), for the \twenty versus \seventy cells, the \twenty versus \fifty cells, and the \fifty versus \seventy cells. 
The smaller of each simulation cell pairing is the template that is translated in space within the larger cell.
The cross-correlation amplitude provides a relative comparison of the similarity between dislocation patterns in the three simulation cells.
Thus, this approach examines whether the smaller simulation cell can act as a valid RVE for reproducing the dislocation structure predicted by the larger simulation cell.
The periodic image of the larger cell was incorporated in this analysis to avoid any artificial influence near the boundaries.
Figures\,\ref{fig:microstructure_crosscorrelation}(a)-(c) show the contour plots of the amplitude of the normalized 3D cross-correlation function for the TEM-like foils shown in Fig.\,\ref{fig:5_Dislocation_microstructure_periodic}.
Only isolated and clearly delimited peaks in the amplitude of the 3D cross-correlation function are observed in the \twenty - \seventy cell pairing, as shown in Fig.\,\ref{fig:microstructure_crosscorrelation}(a).
A more uniform distribution of the amplitudes can be observed in the \twenty - \fifty cell pairing (Fig.\,\ref{fig:microstructure_crosscorrelation}(b)) and more clearly in the \fifty - \seventy cell pairing (Fig.\,\ref{fig:microstructure_crosscorrelation}(c)).
For the latter, no dominant shape can be identified and the distribution of the correlation amplitude qualitatively resembles a natural, random heterogeneous dislocation structure, such as in the autocorrelation function of the \seventy simulation cell. 

%
%

The radial average of the 3D cross-correlation function for each pairing is shown in Fig.\,\ref{fig:microstructure_crosscorrelation}(d).
On average, the cross-correlation amplitude is almost one order of magnitude higher in the \fifty - \seventy cell pairing as compared to the \twenty - \seventy cell pairing.
This indicates that the dislocation structure predicted by the \fifty simulation cell is more closely resembling that predicted by the \seventy cell as compared to that predicted by the \twenty cell.
These results show a clear influence of the periodic simulation cell size on adequately predicting the heterogeneous dislocation structure observed in the reference \seventy simulation cell.

\subsection{Characterizing the variations in the dislocation fluxes during the cool-down simulations}

\begin{figure}[h!]
    \centering
    \includegraphics[width=\textwidth]{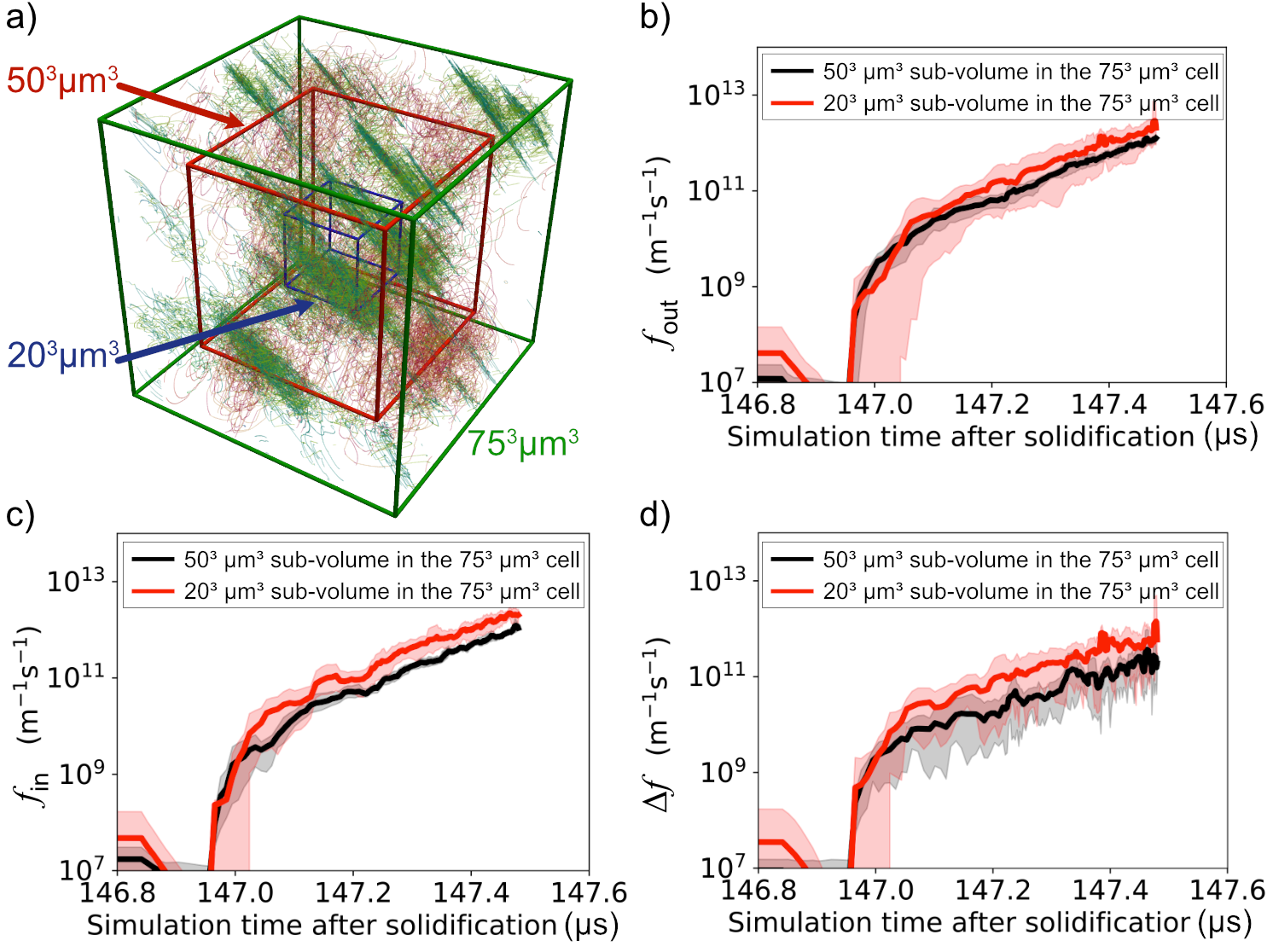}
    \caption{(a) Schematics of two cubic \twenty and \fifty sub-volumes positioned at the center of the \seventy simulation cell. The evolution of the dislocation: (b) flux-out ($f_\mathrm{out}$); and (c) flux-in ($f_\mathrm{in}$) versus time as computed for 27 different positions of the \twenty and \fifty sub-volumes. (d) The dislocation flux difference,  $\Delta f = |f_\mathrm{out} - f_\mathrm{in}|$, versus time. In (b)-(d), the solid lines represents the average over the 27 different positions, while the shaded regions represent the $10^\mathrm{th}$ and $90^\mathrm{th}$ percentile range.} 
    \label{fig:7_Dislocation_flux}
\end{figure}

The \seventy volume simulations cell is subdivided into multiple \twenty and \fifty \textit{sub-volumes} (see details in Section \ref{sec:Method_flux}). 
Here, we use the term ``sub-volume'' to refer to a section of the dislocation structure that was extracted from the reference simulation cell.
The dislocation flux across the boundaries of these sub-volumes is then computed with the goal of identifying a size-invariant periodic cell where the in-flux and out-flux are equal.
Fig.\,\ref{fig:7_Dislocation_flux}(a) shows the 3D dislocation structure after $0.0062\%$ von Mises plastic strain as predicted by 3D DDD simulations using the \seventy simulation cell. The wireframe of a \twenty and \fifty sub-volumes positioned at the center of the simulation cell are also shown. In the following calculations the position of each sub-volume size is varied 27 times by translating each sub-volume from the center of the \seventy simulation cell by increments of $\pm 10\,\upmu$m in the $X$, $Y$, and $Z$ directions, respectively. The dislocation density flux, averaged over all slip systems and from the six surfaces of each sub-volume, is then computed every 2500 time-steps (i.e., a total of 166 time steps). 
The evolution of both the dislocation flux-out and flux-in (see Section \ref{sec:Method_flux} for details) for each sub-volume versus time, are shown in Fig.\,\ref{fig:7_Dislocation_flux}(b) and (c), respectively.
The difference between the flux-out and the flux-in is also shown in Fig.\,\ref{fig:7_Dislocation_flux}(d).
The shaded regions show the $10^\mathrm{th}$ and $90^\mathrm{th}$ percentile confidence range for all 27 different positions of the cubic sub-volumes, while the solid lines show the respective average.
In all cases, an increase in the absolute dislocation flux and in the dislocation flux difference can be observed.
The average flux-in and flux-out (shown by the solid lines) are almost the same for the \twenty and the \fifty sub-volumes.
However, a higher variance (the difference between the $10^\mathrm{th}$ and $90^\mathrm{th}$ percentile) in dislocation flux across the 27 positions is evident for the \twenty sub-volume as compared to the \fifty sub-volume.

%
%

\begin{figure}[h]
    \centering
    \includegraphics[width=\textwidth]{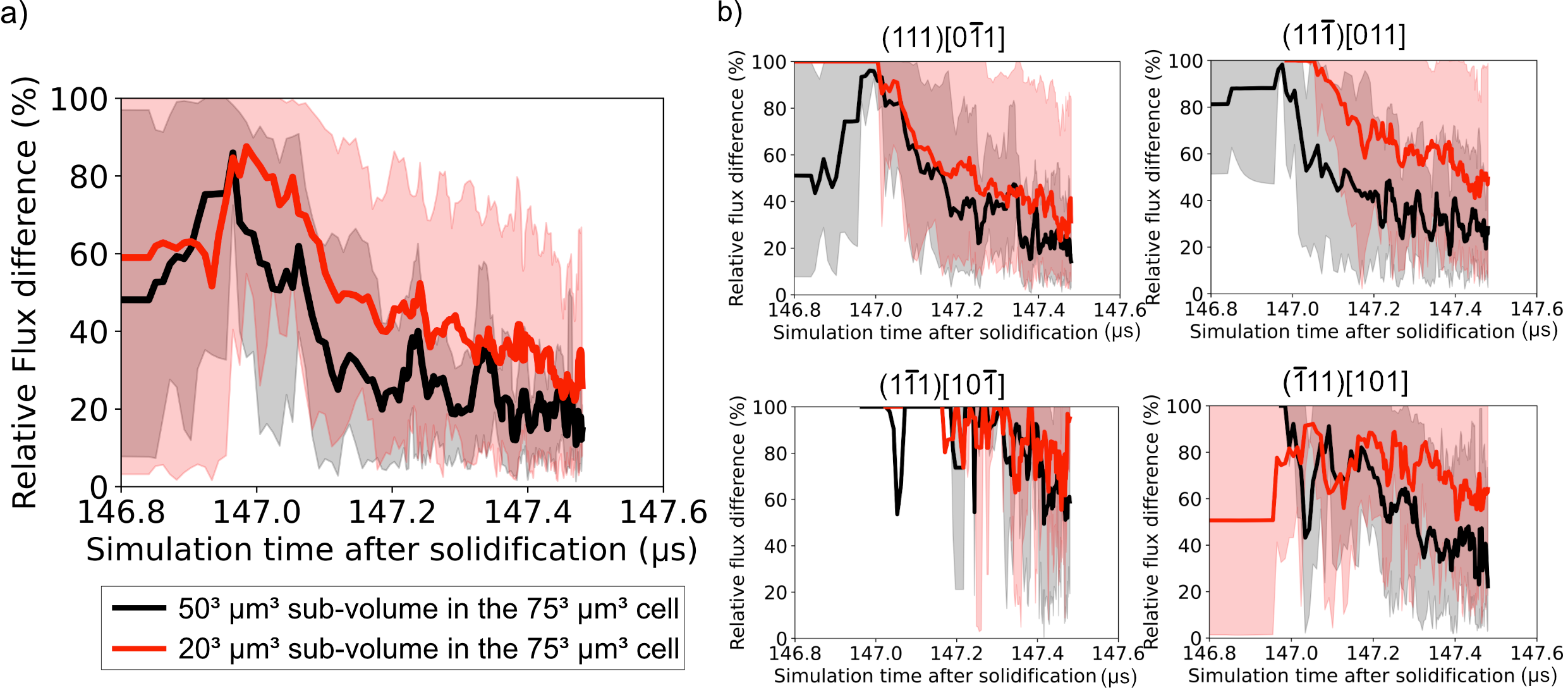}
    \caption{(a) Normalized dislocation flux difference ($\Delta \bar{f} = |\Delta f|/\mathrm{max}\langle f_\mathrm{out},f_\mathrm{in}\rangle$) for the \twenty and \fifty sub-volume sizes at different positions inside the \seventy simulation cell as a function of simulation time. The shaded region represents the $10^\mathrm{th}$ and $90^\mathrm{th}$ percentile confidence range for the 27 different positions. (b) Normalized dislocation flux differences for four slip systems (one of each slip system bundle shown in Fig.\,\ref{fig:3_Density_evolution}(c)). }
    \label{fig:8_Delta_Flux}
\end{figure}

In order to provide a more comparative and generalized measure of the heterogeneity in the dislocation structure formation, the dislocation flux difference normalized by the maximum flux leaving or entering the sub-volume (see Section \ref{sec:Method_flux} for details) is shown in Fig.\,\ref{fig:8_Delta_Flux}.
This normalized flux difference is a measure that relates the local heterogeneity in dislocation flux to the overall dislocation activity in the simulation cell and therefore allows for a relative assessment of the observed dislocation flux difference.
A decrease in relative flux difference with increasing simulation time is observed for both sub-volumes starting from up to 80\% to about 20\% relative difference at the end of the simulation.
Throughout the simulation, the averaged value over all 27 positions in the case of the \twenty sub-volume as well as the variance (the difference range between the $10^\mathrm{th}$ and $90^\mathrm{th}$ percentile) exceeds the corresponding measure from the \fifty sub-volume.
When considering different slip systems separately, as shown in Fig.\,\ref{fig:8_Delta_Flux}(b), the effect of the sub-volume size is also observed and even more pronounced for some slip systems.
Here, a relative flux difference of zero for all positions of the averaging volume means the corresponding sub-volume would be large enough to adequately represent the overall dislocation structure evolution predicted by the reference volume for the given point in time. 
The decreasing normalized flux difference for all cases indicates that while the net difference in dislocation flux increases with increasing dislocation density (Fig.\,\ref{fig:7_Dislocation_flux}(d)), its influence on plastic flow decreases when compared to the overall dislocation activity.

%
%

\section{Discussion}
\label{4_Disc}

\subsection{Transient residual stresses and the constraints on choosing an initial dislocation structure}

The constraint in contraction of the re-solidified material at the solid-liquid interface after single-track laser scan results in a change in the load path during cool-down. This is evident by the evolution of the residual stresses and derived nominal Schmid-factors during the cool-down stage showing a transition in the dominant Schmid-factor occurs at about 1100K (Fig.\,\ref{fig:1_FEM}). Such changes in the loading path during cool-down imposes further complexities to DDD simulations as compared to simulations of pure uniaxial load. The DDD simulations here must be able to accurately predict the change in dislocation evolution associated with the changes in the loading path.
The dislocation structure predicted in later stages of cool-down should also be independent of statistical fluctuations in the chosen initial simulation conditions and resulting dislocation structure formation during early cool-down.
This includes the dislocation evolution on slip systems with low and medium nominal Schmid-factors as they become dominant during progressing cool-down (Fig.\,\ref{fig:1_FEM}(b)).

Statistical fluctuations in randomly generated dislocation structures significantly constrain the choice of adequate system volume sizes depending on the dislocation density.
This is evident from the statistical analysis of 900 random initial dislocation structures using different initial dislocation densities and simulation cell sizes, as shown in supplementary Fig.\,S2. 
It is demonstrated that statistical fluctuations can be large in absolute numbers even when averaged over a large volume. 
As a result, a heterogeneous dislocation structure formation could be expected even from small relative changes in dislocation density.
This indicates that the evolution of the dislocation structure and subsequent material properties should not be evaluated based on averaged values alone. 
Instead, it should include further criteria that, e.g., would allow the consideration of the spatial extent of heterogeneous dislocation structure formation.

\subsection{Interplay between dislocation structure formation, dislocation interactions, and strain hardening}

Comparing the evolution of the dislocation density with simulation time in the $20^3\,\upmu$m$^3$, $50^3\,\upmu$m$^3$, and \seventy simulation cells, shown in Fig.\,\ref{fig:3_Density_evolution}, a reduced activity on slip systems having an intermediate nominal Schmid-factor (around 0.25) is observed for the smallest volume.
From the statistical analysis on randomly generated dislocation structures in Supplementary Section S2, it can be inferred that this behavior is caused by the difference in the initial random dislocation structure in each simulation cell.
In contrast, a dislocation evolution in agreement with the slip system activity expected from the Schmid-factors is observed for the \fifty and the \seventy sub-volume simulation cells.

Furthermore, the higher density of dislocation junctions for the same level of plastic strain observed in the \seventy simulation cell (Fig.\,\ref{fig:3_Density_evolution}(d)) indicates more pronounced dislocation interactions on non-coplanar and non-cross-slip systems. 
Based on crystallographically expected junction formations for the different FCC slip system combinations, it can be inferred that slip systems exhibiting an intermediate nominal Schmid-factor (which are inactive in the \twenty simulation cell case) provide a substantial contribution to the formation of a strong dislocation network through formation of Lomer and glissile junctions.
Thus, in addition to a higher density of dislocation junctions predicted by using the \seventy simulation cell, it can be expected that the junctions in both the \seventy and the \fifty sub-volume simulations consist of more stable Lomer reactions. 
With increasing strain and dislocation density, a higher fraction of glissile reactions will also lead to an increase in dislocation density on other slip systems, as shown in Fig.\,\ref{fig:3_Density_evolution}(b) and (c). 

The results indicate that the evolved dislocation structure predicted by the \twenty simulation cell will consist of a less stable dislocation network as compared to that in the \fifty and \seventy simulation cells.
This consequently would result in underestimating the strain hardening after the load path change during progressing cool-down. 
This is in line with previous studies that show that glissile junctions can significantly influence dislocation structure formation and hardening in certain loading orientations \cite{strickerDislocationMultiplicationMechanisms2015a, sudmannsDislocationMultiplicationCrossslip2019}.
Furthermore, an insufficient prediction of dislocation density increase and formation of dislocation structure on less active slip systems might also alter the prediction of cellular dislocation structures with ongoing cool-down as observed in experiments and in meoscale simulations of LPBF AISI 316L stainless steels \cite{sudmannsInterplayLocalChemistry2022,bertschOriginDislocationStructures2020,wangAdditivelyManufacturedHierarchical2018}.

\subsection{Simulation cell size effects}

By arranging the dislocation structure periodically as shown in Fig.\,\ref{fig:4_Dislocation_microstructure_3D}, Fig.\,\ref{fig:5_Dislocation_microstructure_periodic}, and in supplementary Figs.\,S5-S8, a very regular dislocation pattern formation is observed in the \twenty simulation cell that is not observed in the larger simulation cell cases.
This is also evident from the autocorrelation analyses of all three simulation cell sizes, as shown in Fig.\,\ref{fig:microstructure_autocorrelation}.
No periodicity is evident in the autocorrelation functions for the two larger simulation cells, which indicates a transition towards a more random and natural dislocation structure formation that is increasingly independent from the prescribed boundary conditions.

The low number of active slip systems in the \twenty cell favors the formation of periodically extending (i.e. technically infinite) regular dislocation cells with a spacing that is dominated by the periodic dimension of the DDD simulation box. 
This dislocation structure formation is characterized by intense cross-slip at very high temperatures.
Therefore, the aspects regarding treatment of periodic boundaries in DDD simulations, such as the kinematic consistency of dislocation lines or spurious annihilation of dislocations interacting with their own periodic image as discussed e.g. in \cite{pachaury2022discrete,bulatov2000periodic,madec2004use}, are unlikely to be of concern for the given set of simulations.
Instead, the observed simulation cell size effect can be attributed to the fact that the dislocation structure formation is driven by single dislocation walls which do not interact with dislocation structures that were formed on other slip systems before encountering a periodic boundary condition.
Subsequently, dislocation walls are formed with a spacing that reflects the size of the primary cell due to intense cross-slip at high temperatures. 
In contrast, the dislocation walls in the \fifty and the \seventy simulation cells are limited in their spatial extent for the same level of plastic strain.
In this case, the dislocations encounter other structures that were formed within the volume before reaching the simulation cell boundaries, and therefore form a more realistic and stable dislocation network.
Cross-correlating the voxel-averaged dislocation density structures between the investigated simulation cell sizes demonstrates that the dislocation structure predicted by the smallest cell is not representative of a more realistic dislocation structure such as that predicted by the largest cells.
This can be inferred from the fact that only isolated peaks in the correlation function are visible by pairing the \twenty cell with the \seventy cell, whereas a more uniform distribution of the correlation function is observed by pairing the \fifty cell with the \seventy cell.

\subsection{The representative volume elements in discrete dislocation dynamics simulations}

In all 3D DDD simulations we used periodic boundary conditions to mimic the dislocation structure evolution in a bulk single-crystalline material subjected to LPBF 316L SS processing. 
This approach is based on the idea of ``Representative Volume Elements'' (RVEs), where each RVE is assumed to be large enough to incorporate all physics of the computational representation of the bulk material, but is as small as possible to limit the computational effort.
To identify the optimal RVE size in 3D DDD simulations of cool-down during LPBF 316L SS processing, we calculated the dislocation fluxes through \textit{sub-volumes} of different sizes positioned within the \seventy reference cell.
Thereby, we assessed if smaller sub-volumes are capable of reproducing the dislocation structure evolution observed in a bulk volume (which is defined here as the largest simulation cell studied).
If this is the case, the dislocation structure evolution can be considered invariant of the size and presence of computational boundaries, hence, the sub-volume can serve as a \textit{RVE} of the single crystal bulk response as predicted by the DDD simulation with the largest volume size.
The net difference between the dislocation flux-in and flux-out of the sub-volumes normalized by the total flux (Fig.\,\ref{fig:8_Delta_Flux}(a)) is used as a measure to capture the heterogeneity in dislocation plasticity in the simulation volume.
A significant spatial variance of the dislocation flux averaged over the boundaries of the sub-volumes positioned within the \seventy reference cell is observed throughout most of the simulation time. 
It can thus be concluded that the heterogeneous dislocation structure observed later in the \seventy simulation volume is formed very early in the simulation.

The investigation presented here is strongly influenced by the specific load cycle observed in LPBF processing of AISI 316L. 
A more general definition of RVEs for DDD would have to consider further characteristics depending on initial dislocation structures, crystal orientations, strain rates, or other loading scenarios that might introduce completely different challenges in terms of structure formation and heterogeneity.
However, various conclusions can be transferred to other scenarios.
For example, the results show that while the net dislocation flux difference increases with simulation time, the normalized flux difference decreases over time.
This means that the relevance of single events for plastic flow decreases in relation to the overall dislocation activity, which in turn increases due to an increasing dislocation density.
It can thus be concluded that the evolution of the dislocation structure gradually evolves towards homogeneous plastic flow during progressing cool-down.
This is especially true in the case when choosing a \fifty sub-volume, which approaches size-invariant periodicity with ongoing simulation.

\section{Conclusions}

Using the example of single track LPBF scans of AISI 316L stainless steels, we addressed the relationship between micro-scale dislocation structure evolution predicted by 3D DDD simulations using different simulation volume sizes and macro-scale thermo-mechanical stresses predicted by finite element simulations.
The results show that an insufficient size of periodic simulation volumes can result in dislocation patterns that reflect the boundaries of the volume.
A less pronounced dislocation interaction is observed for smaller simulation volume sizes for the same dislocation density and plastic strain, indicating an influence of simulation volume sizes on strain hardening.
Thus, the change in load path observed during cool-down in the present study of LPBF AISI 316L indicates that an accurate representation of dislocation processes on non-dominant slip systems is essential for an accurate prediction of mechanical properties. 

Furthermore, we characterized and identified representative volume elements in DDD simulations by a dislocation flux condition that captures the heterogeneity in dislocation plasticity in a bulk volume by averaging over sub-volumes that are positioned within the bulk volume.
A heterogeneity in dislocation flux is observed, which should be accounted for by existing approaches of modeling bulk material behavior in DDD and other related simulation methods when aiming at replicating bulk material behavior with periodic boundary conditions.
As such, the history of the deformation process has to be considered when choosing appropriate simulation volume sizes in DDD simulations.

Although we applied the method to LPBF AISI 316L stainless steels, the conclusions can be transferred to other applications involving a change in load path.
On the macroscopic level, this could include, e.g., metal forming processes or bi-directional cyclic loading.
One example for a load path change on the microscopic level is the residual stress formation in surface treatment processes.

Several opportunities for future research arise from these conclusions.
Although it has been shown that avoiding artificial dislocation patterns emerging from very low dislocation densities requires very large simulation volumes, modeling the evolution of dislocation structures at higher densities using these simulation volume sizes is rather impractical.
There are several conceivable ways to balance the trade-off between a desired system size invariance and computational cost.
First, instead of enforcing strict line continuity across periodic boundaries, an additional dislocation flux could be added originating from virtual dislocation sources outside the system boundaries. 
The dislocation flux necessary to ensure a dislocation structure evolution equivalent to larger RVEs could be estimated, e.g., from the plastic slip predicted by coarse-grained crystal plasticity or dislocation-based plasticity simulations.
Second, it might be plausible to dynamically adjust the system volume size to the required dimension, where the latter could be predicted by a dislocation flux condition like the one proposed here.
Thereby, this work provides the basis for a physically-based and reliable description of representative volume elements for DDD simulations that can be used to characterize heterogeneous dislocation microstructure evolution and mechanical properties in complex loading scenarios.
Further, the proposed methods of characterizing the dislocation structure evolution could help to provide crystal plasticity scale models with guidance on up-scaling of meso-scale simulations that would enable accurate CP-FEM simulations of AM printed parts.

\section*{Acknowledgements}

The authors gratefully acknowledge support for this work from the Office of Naval Research through the Naval Research Laboratory’s core funding and grant number N00014-18-1-2858 as well as by the U.S. National Science Foundation (NSF) CAREER award CMMI-1454072.
Some simulations were conducted at the Advanced Research Computing at Hopkins (ARCH) core facility (rockfish.jhu.edu), which is supported by an NSF grant number OAC-1920103. Some simulations were also conducted using the Extreme Science and Engineering Discovery Environment (XSEDE) Expanse supercomputer at the San Diego Supercomputer Center (SDSC) through allocation TG-MAT210003. XSEDE is supported by National Science Foundation grant number ACI-1548562.

\section*{Competing interests}

The Authors declare no Competing Financial or Non-Financial Interests

\section*{Data Availability}

All data are available from the corresponding authors upon reasonable request.

\section*{Author contributions}

\textbf{Jaafar El-Awady and John Michopoulos} designed and led the computational aspects of this project.
\textbf{Athanasios Iliopoulos, Andrew Birnbaum, and John Michopoulos} developed, conducted and evaluated the FEM simulations.
\textbf{Markus Sudmanns and Jaafar El-Awady} developed and evaluated the 3D DDD simulations.
All authors contributed to the analysis and discussion of the results.
\textbf{Markus Sudmanns and Jaafar El-Awady} wrote the initial manuscript and all co-authors provided input to the final manuscript.

\appendix

\section{Correlation analyses of dislocation microstructures}
\label{sec:Correlation}

Three-dimensional correlation analyses of the voxel-averaged dislocation density in the $20^3\,\upmu$m$^3$, $50^3\,\upmu$m$^3$, and $75^3\,\upmu$m$^3$ simulation cells are used to provide an identification and quantification of similarities in the predicted dislocation structures.
To prepare the data for the analysis, we calculated local dislocation densities by discretizing the full simulation volume into voxels having edge length of $1\,\upmu$m$^3$. This creates system volumes consisting of $20^3$, $50^3$, and $75^3$ voxels for the $20^3\,\upmu$m$^3$, $50^3\,\upmu$m$^3$, and $75^3\,\upmu$m$^3$ simulation cells, respectively.

Using the generated datasets, we first calculated 3D spatial autocorrelation functions of the local dislocation densities in all simulations to identify potential periodic patterns in the dislocation microstructure. 
The autocorrelation describes the correlation between a dataset and a copy of itself, which is translated in space by a vector $\mathbf{u}$.
This technique is used in digital image processing to correlate one image with itself \cite{jahne2005digital, HEILBRONNER1992351} and was also applied to analyses of dislocation microstructures \cite{sudmannsInterplayLocalChemistry2022,robertsonDigitalRepresentationQuantification2021}.  
For an efficient calculation of the autocorrelation function, we make use of the relationship between the autocorrelation function and the Fourier transform, known as the Wiener-Khinchin theorem \cite{wiener1930generalized, khintchine1934korrelationstheorie}.
This theorem states that the autocorrelation function of a stationary signal is equal to the Fourier transform of the power spectral density.
The autocorrelation function $C_\mathrm{auto}(\mathbf{u})$ of a signal $F$ (the generated dataset in case of our analysis) is then given by
\begin{equation}
C_\mathrm{auto}(\mathbf{u}) = \frac{E^{-1}\left[E(F)\bar{E}(F)\right]}{\sqrt{\sum F^2\sum F^2}}, 
\label{eq:autocorrelation}
\end{equation}
where $E(\cdot)$ is the Fourier transform, $E^{-1}$ its inverse, and $\bar{E}(\cdot)$ the complex conjugate.
The function is normalized such that $C_\mathrm{auto}(\mathbf{u})$ is within the interval $[0,1]$, where a correlation of $1$ represents the identity, i.e. a zero translation in space.

In a second step, we compared the similarities between the microstructure formation in the $20^3\,\upmu$m$^3$, $50^3\,\upmu$m$^3$, and $75^3\,\upmu$m$^3$ simulation cells by pairwise cross-correlation.
We focused on the spacial extent and periodicity of heterogeneous dislocation structures such as dislocation cells or walls.
Therefore, we adopted the cross-correlation technique from signal processing, which is used to to identify the similarity between two periodic signals \cite{yoo2009fast}.
In the same way it can be used in digital image processing for template-image matching to find smaller parts of an image within a larger image \cite{briechle2001template}.
For our purposes, we describe the cross-correlation by the correlation function $C_\mathrm{cross}(\mathbf{u})$ between two datasets $F(x,y,z)$ and $G(x-u,y-v,z-w)$, where $G$ serves as a template which is translated in space described by the vector $\mathbf{u} = [u,v,w]$.
The correlation coefficient can then be derived as 
\begin{equation}
C_\mathrm{cross}(\mathbf{u}) = \frac{\sum_{x,y,z}\left(F(x,y,z)G(x-u,y-v,z-w)\right)}{\sqrt{\sum_{x,y,z}F(x,y,z)^2\sum_{x,y,z}G(x-u,y-v,z-w)^2}}, 
\label{eq:correlation}
\end{equation}
which is adopted from \cite{briechle2001template,yoo2009fast}. 
We note that contrary to the often employed approach in digital image processing, the dataset as predicted by the DDD simulations is not zero-normalized in order to allow for a more physical interpretation of the resulting correlation coefficients $C_\mathrm{cross}(\mathbf{u})$ in the range of $[0,1]$.
We further note that the autocorrelation given by Eq.\,(\ref{eq:autocorrelation}) is a special case of the cross-correlation, for which we set $F$ = $G$ in Eq.\,(\ref{eq:correlation}).
Applying Eq.\,(\ref{eq:correlation}) to our 3D DDD simulations, the template $G$, which is translated in space by the vector $\mathbf{u}$, corresponds to the \twenty and \fifty simulation cells, respectively.
The templates are correlated with the respective larger \fifty and \seventy simulation cells, where we include the periodic images of the larger simulation cell of each pairing to correctly capture the periodicity of the microstructure.





\section*{References} 
\bibliographystyle{unsrt}

\end{document}







\maketitle

\renewcommand{\thesection}{S \arabic{section}}
\renewcommand{\thesubsection}{S 1.\arabic{subsection}}
\renewcommand{\thetable}{S\arabic{table}}
\renewcommand{\theequation}{S \arabic{equation}}
\setcounter{table}{0}

\renewcommand{\thefigure}{S\arabic{figure}}
\setcounter{figure}{0}

\section{Thermo-elaso-plastic Finite Element Model}
The thermo-elaso-plastic finite element model (FEM) is based on a Lagrangian approach which couples structural mechanics and heat transfer physics. 
The equation of motion in the absence of inertial and body forces is:
\begin{equation}
	 \nabla \cdot \left( \mathbf{F} \mathbf{S} \right)^T = \mathbf{0}, 
\end{equation}
where  $\mathbf{F} = \mathbf{I} + \nabla \mathbf{v}$ is the deformation gradient tensor, $\mathbf{v}$ is the displacement vector, and $\mathbf{S}$ is the Second Piola-Kirchhoff stress tensor. The multiplicative decomposition of the deformation gradient is:
\begin{equation}\label{eq:defgrad}
	\mathbf{F} = \mathbf{F}_\mathrm{el}\mathbf{F}_\mathrm{inel} = \mathbf{F}_\mathrm{el}\mathbf{F}_\mathrm{pl}\mathbf{F}_\mathrm{th},
\end{equation}
where the index $()_\mathrm{el}$ denotes the elastic strain, $()_\mathrm{pl}$ the plastic strain, and $()_\mathrm{th}$ the strain due to thermal expansion.
The relationship between the Green-Lagrange strain tensor of type $X$ with the deformation gradient is given by ${\bm{\varepsilon}}_X = \nicefrac{1}{2}\left( \mathbf{F}_X^T \mathbf{F}_X - \mathbf{I} \right)$. 
The infinitesimal strain tensor can be written by additive decomposition:
\begin{equation}
	\bm{\varepsilon} = \bm{\varepsilon}_\mathrm{el} + \bm{\varepsilon}_\mathrm{th} + \bm{\varepsilon}_\mathrm{pl},
\end{equation}
where $\varepsilon_{th_{ii}}=\alpha \left( T - T_\mathrm{ref} \right)$ and  $\varepsilon_{th_{ij}}= 0 , i,j=\left\{x,y,z\right\}, i \neq j$, $\alpha$ is the coefficient of thermal expansion $\alpha$, and $T_\mathrm{ref}$ is the reference temperature denoting the temperature at which the thermal strain vanishes (i.e., $\bm{\varepsilon}_\mathrm{th}=\mathbf{0}$).

The numerical implementation of the material model involves the classical trial stress approach \cite{owen1980finite}. 
To account for the resolidification of the melted material inside the melt pool, our implementation uses the following trial strain form:
\begin{equation}\label{eq:trialstrain}
	\bm{\varepsilon}_\mathrm{trial} = \bm{\varepsilon} - \bm{\varepsilon}_\mathrm{pl} - \bm{\varepsilon}_\mathrm{th} - \mathbf{e}_\mathrm{resol},
\end{equation}
where $\mathbf{e}_\mathrm{resol}=\bm{\varepsilon}_\mathrm{resol}-\bm{\varepsilon}_\mathrm{th_{resol}}$ is the elastic strain at resolidification, with $\bm{\varepsilon}_\mathrm{resol}$ being the total strain and $\bm{\varepsilon}_\mathrm{th_{resol}}$ the thermal strain at resolidification. It should be noted that the plastic strain is zeroed when the temperature exceeds the melting temperature. The reference temperature $T_\mathrm{ref}$ is changed to the melting temperature since thermal stress will increase with progressing cool-down.
It can be easily shown that the subtraction of the thermal expansion term in Eq.\,\eqref{eq:trialstrain} is equivalent to changing the reference temperature to the melting temperature: 
\begin{equation}  
	\begin{split}
	&-\bm{\varepsilon}_\mathrm{th}-\left( -\bm{\varepsilon}_\mathrm{th_{resol}} \right) = -\left( \bm{\varepsilon}_\mathrm{th}- \bm{\varepsilon}_\mathrm{th_{resol}} \right) = \\ & = - \left( \alpha \left(T -T_\mathrm{ref} \right)  - \alpha \left(T_\mathrm{melt}-T_\mathrm{ref} \right) \right)=\\
	& = - \alpha \left( T -T_\mathrm{ref} -T_\mathrm{melt} + T_\mathrm{ref} \right)  = \\
	& - \alpha \left( T -T_\mathrm{melt} \right) 
	\end{split} 
\end{equation}

The local form of the energy conservation takes the form of the heat transfer equation in the material frame  expressed by:
\begin{equation}
	\rho\left( T \right) C_p\left( T \right) \frac{\partial T}{\partial t}  + \nabla \cdot \mathbf{q} = Q_\mathrm{ted},
\end{equation}
where $T$ is the temperature, $\mathbf{q}=-k\left( T \right) \nabla T$ is the heat flux vector defined by the Fourier constitutive equation, $\rho\left( T \right)$ is the temperature dependent material density, $C_p\left( T \right)$ is the temperature dependent heat capacity, $k\left( T \right)$ is the temperature dependent thermal conductivity, and $Q_\mathrm{ted} = \alpha T : \frac{\partial S}{\partial t}$ is the thermoelastic damping.

\begin{figure}[tb]
	\includegraphics[width=\linewidth]{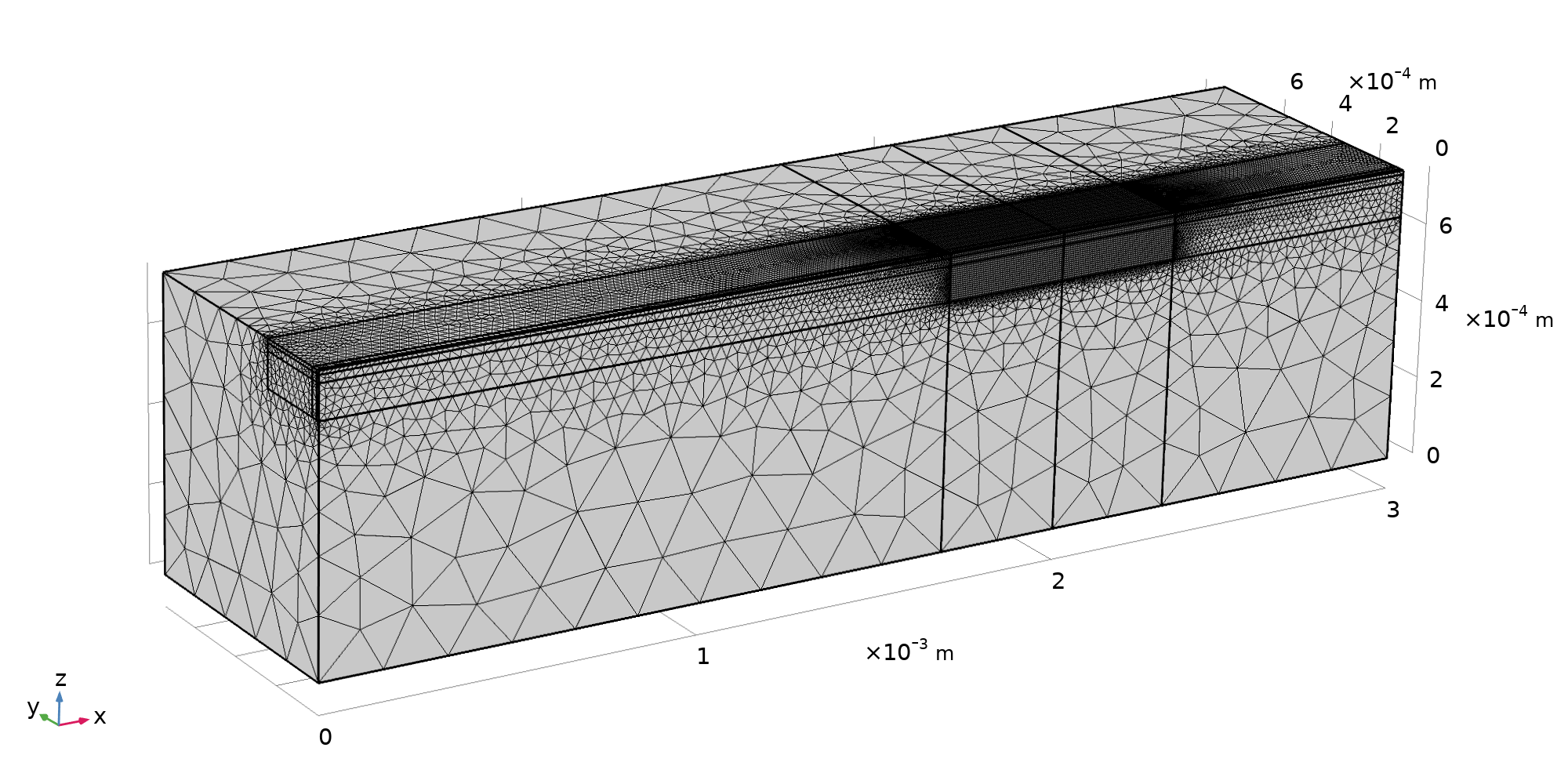}
	\caption{The meshed domain used in the finite element simulations. The simulation results were probed at the locally refined mesh at $x = 2\times10^{-3}$m.}
	\label{fig:mesh}
\end{figure}

The discretized domain used in the simulation is shown in Fig. \ref{fig:mesh} and consists of approximately 620,000 elements.

\section{Statistics of Initial dislocation microstructures for different sub-volume sizes}
The DDD simulations are initialized by randomly generating dipolar dislocations, each having random position, length, and orientation.
Given that the initial dislocation density is relatively low, the influence of simulation cell size on the initially generated dislocation structure must be characterized.
As such, 100 random realizations of the initial dipolar dislocation structures are generated within a $20^3\,\upmu$m$^3$, $50^3\,\upmu$m$^3$, and \seventy simulation cell, respectively. 
The dislocation densities used are $\rho_0 = 10^9\,$m$^{-2}$, $10^{10}\,$m$^{-2}$, and $10^{11}$m$^{-2}$ for all simulation cell sizes, creating a set of $9\times100$ initial dislocation structures.

\begin{figure}[h]
    \centering
    \includegraphics[width=\textwidth]{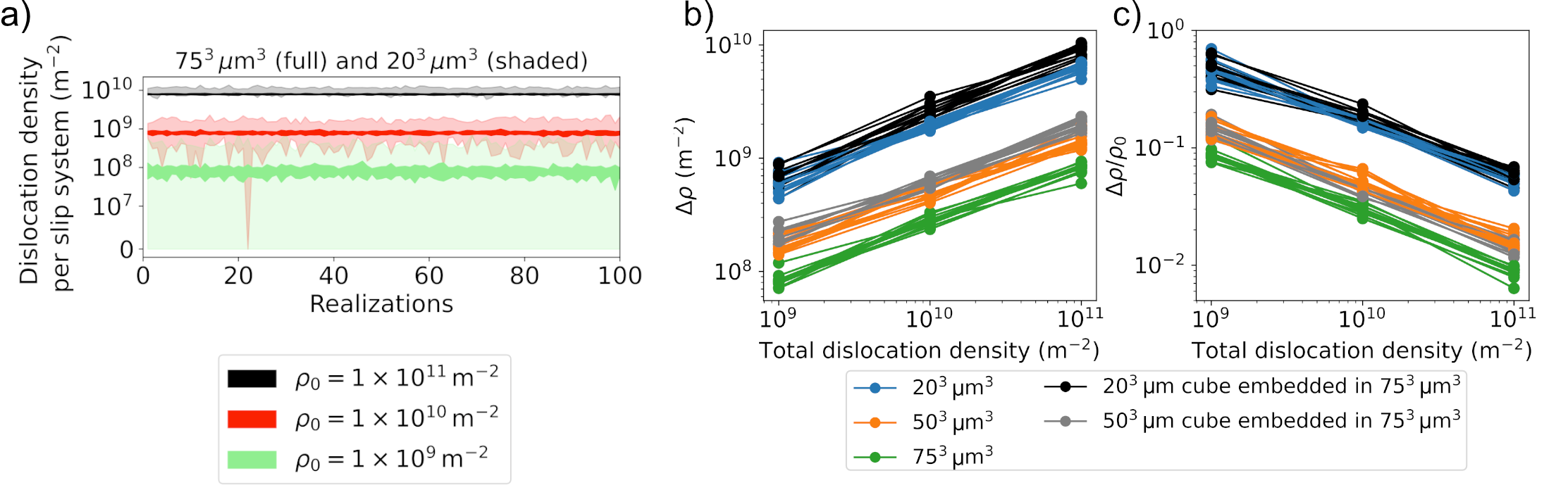}
    \caption{(a) Variation in the dislocation density per slip system for different total initial dislocation density $\rho_0$, random realizations of the initial dislocation structure, and different simulation cell sizes. The dark region represents $20^3\,\upmu$m$^3$ simulation cell, while the lighter colored regions represent the \seventy simulation cell. (b) Absolute difference in dislocation density per slip system, $\Delta\rho$, averaged over all random realizations of the initial dislocation structure and averaged over the simulation cell as a function of total average dislocation density for the $20^3\,\upmu$m$^3$, $50^3\,\upmu$m$^3$, and \seventy simulation cells. (c) Shows the data shown from (b) normalized by the respective total dislocation density $\rho_0$. Additionally, the corresponding data averaged over a \twenty and \fifty sub-volume placed inside the \seventy volume is shown in (b) and (c).}
    \label{fig:appendix_initial_density}
\end{figure}

Figure\,\ref{fig:appendix_initial_density}(a) shows the range of generated initial dislocation densities on each slip system as a colored area bounded by the minimum and maximum density in each realization, respectively, for the different average total dislocation densities. 
Here, the darker and lighter colored areas represent the variation of dislocation densities per slip system in the $20^3\,\upmu$m$^3$, and \seventy simulation cell, respectively.
It is observed that for an average dislocation density of $10^{10}\,$m$^{-2}$ and $10^{11}\,$m$^{-2}$, the variations of the dislocation density on different slip systems is small compared to the average dislocation density.
However, in case of a total density of $10^{9}\,$m$^{-2}$ in the \twenty simulation cell, the minimum value in dislocation density per slip system is zero in all random realizations, which means that some slip systems remained unpopulated.

We further quantify the variations in dislocation density per slip system by calculating the difference $\Delta \rho$ between the maximum and the minimum density on each slip system from the 100 realizations for each total density and sub-volume size.
The change in absolute, $\Delta \rho$, and normalized difference with the total dislocation density, $\Delta \rho/\rho_0$, are shown in Fig.\,\ref{fig:appendix_initial_density}(b) and (c), respectively.
A clear influence of simulation cell size and total dislocation density on the variance in dislocation density per slip system is evident, where a decreasing variance is observed for large simulation cell sizes and higher total dislocation densities.
It can be seen that the initial dislocation structure is statistically equivalent when extracting a smaller (i.e. $20^3\,\upmu$m$^3$, or $50^3\,\upmu$m$^3$) sub-volume out of the largest \seventy simulation cell as compared to generating a random initial dislocation structure in simulation cells of the same sizes. 
The results show that while variations in dislocation density might be small relative to the total dislocation density, they can still be large in absolute numbers, especially when averaged over smaller sections of the systems.
This can result in a heterogeneous dislocation structure formation, although a low dislocation density variation might indicate otherwise.

\section{Additional supplementary figures}

\begin{figure}[h]
    \centering
    \includegraphics[width=0.5\textwidth]{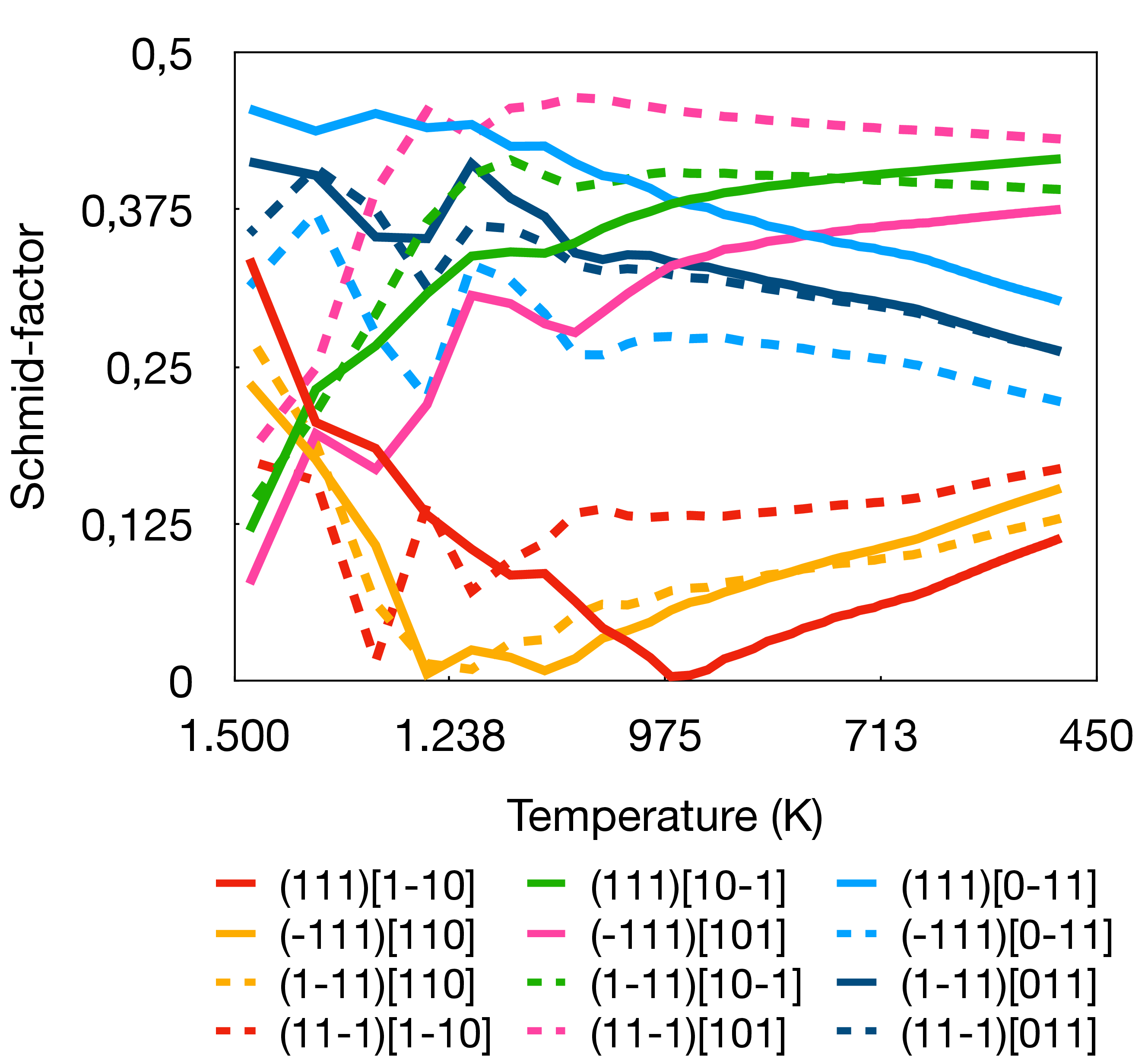}
    \caption{The evolution of the Schmid-factors on all slip systems as a function of temperature. Solid lines and dashed lines having the same color represent a slip system and its corresponding cross-slip system.}
    \label{fig:Schmid-factors_appendix}
\end{figure}

The evolution of the Schmid-factors with temperature for all slip systems is shown in Fig.\,\ref{fig:Schmid-factors_appendix} which have been calculated using the evolution of the stress components predicted by the FEM simulations. For simplicity, in these calculations the $[100]$, $[010]$, and $[001]$ crystal directions are assigned  parallel to the $x$ (scanning), $y$ (normal), and $z$ (build) axes, respectively.

\begin{figure}[h]
    \centering
    \includegraphics[width=\textwidth]{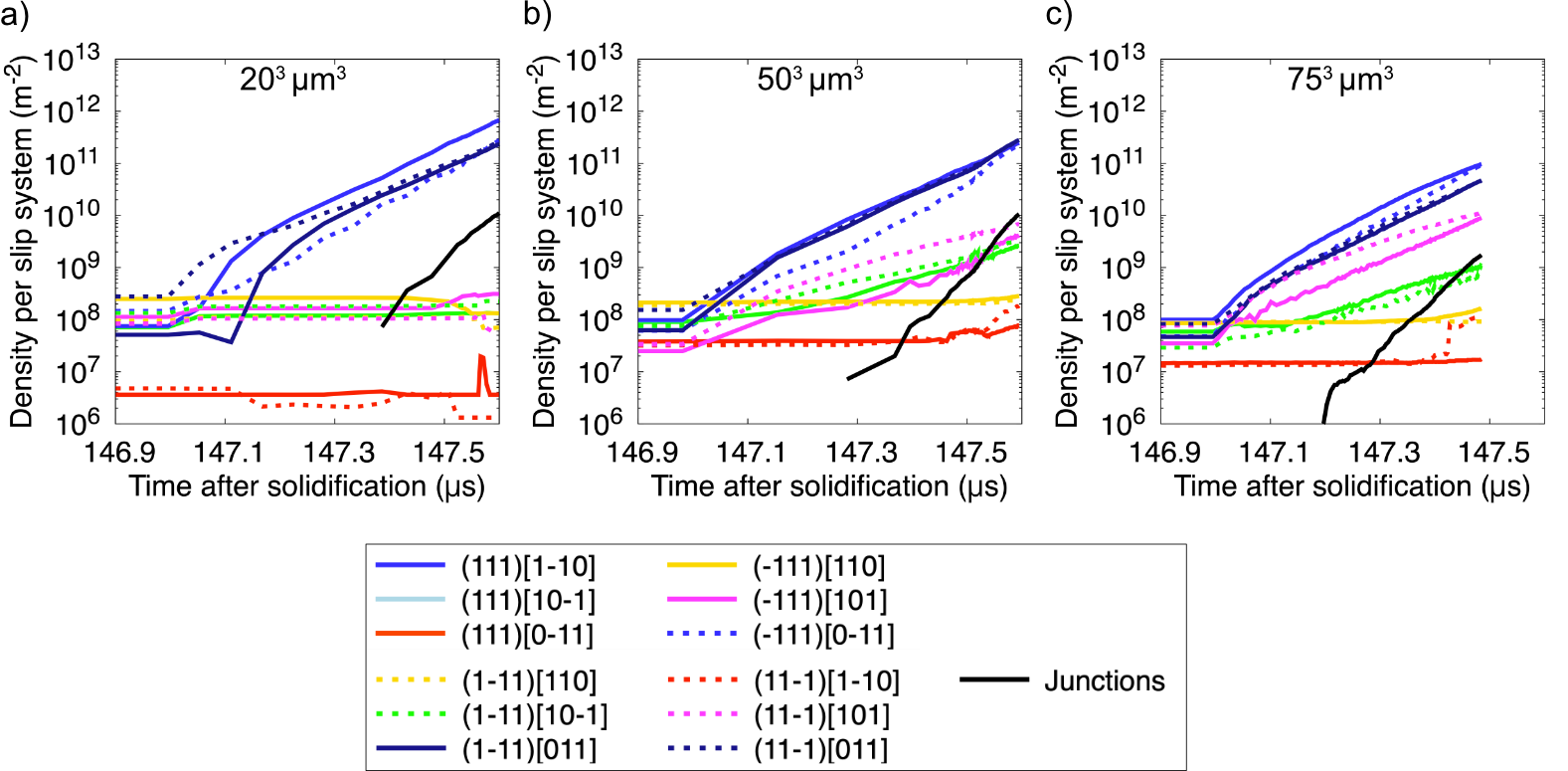}
    \caption{Evolution of the dislocation density versus time on the twelve FCC slip systems  as well as that for junctions for the: (a) $20^3\,\upmu$m$^3$; (b) $50^3\,\upmu$m$^3$; and (c) \seventy simulation cells. The simulation time $t = 0$ represents the start of the DDD simulations, which coincide with the instance of solidification.}
    \label{fig:Schmid-Density_evolution_appendix}
\end{figure}

Figure\,\ref{fig:Schmid-Density_evolution_appendix}(a)-(c) shows the predicted time evolution of the dislocation density for all slip-system for the three  cubic simulation cell sizes of $20^3\,\upmu$m$^3$, $50^3\,\upmu$m$^3$, and $75^3\,\upmu$m$^3$.

\begin{figure}[h]
    \centering
    \includegraphics[width=\textwidth]{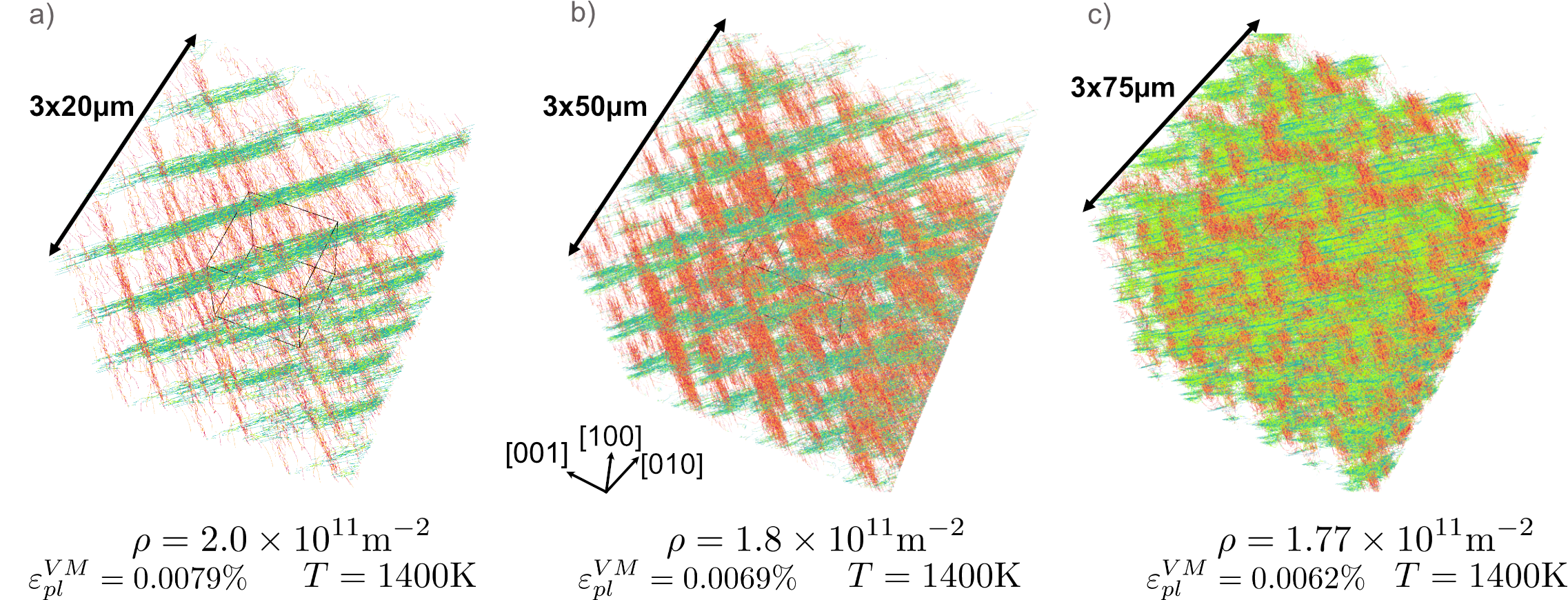}
    \caption{The 3D dislocation structure as predicted by the DDD simulations at 1400K for the (a) $20^3\,\upmu$m$^3$, (b) $50^3\,\upmu$m$^3$, and (c) \seventy simulation cell. Shown are three periodic replicas of the simulation volume in each directions. The original simulation volume is represented by the black cube at the center.}
    \label{fig:appendix_Dislocation_microstructure_all_3d}
\end{figure}

The 3D dislocation structure as predicted by the DDD simulations using the $20^3\,\upmu$m$^3$, $50^3\,\upmu$m$^3$, and \seventy simulation cells are shown in Fig.\,\ref{fig:appendix_Dislocation_microstructure_all_3d}. In all cases three periodic replicas are shown in all three directions (i.e., showing cubic volumes of $60^3\,\mu$m$^3$, $150^3\,\mu$m$^3$, and $225^3\,\mu$m$^3$) to visualize the effect of the periodic boundaries on the formation of extended dislocation walls.

\begin{figure}[h!]
    \centering
    \includegraphics[width=\textwidth]{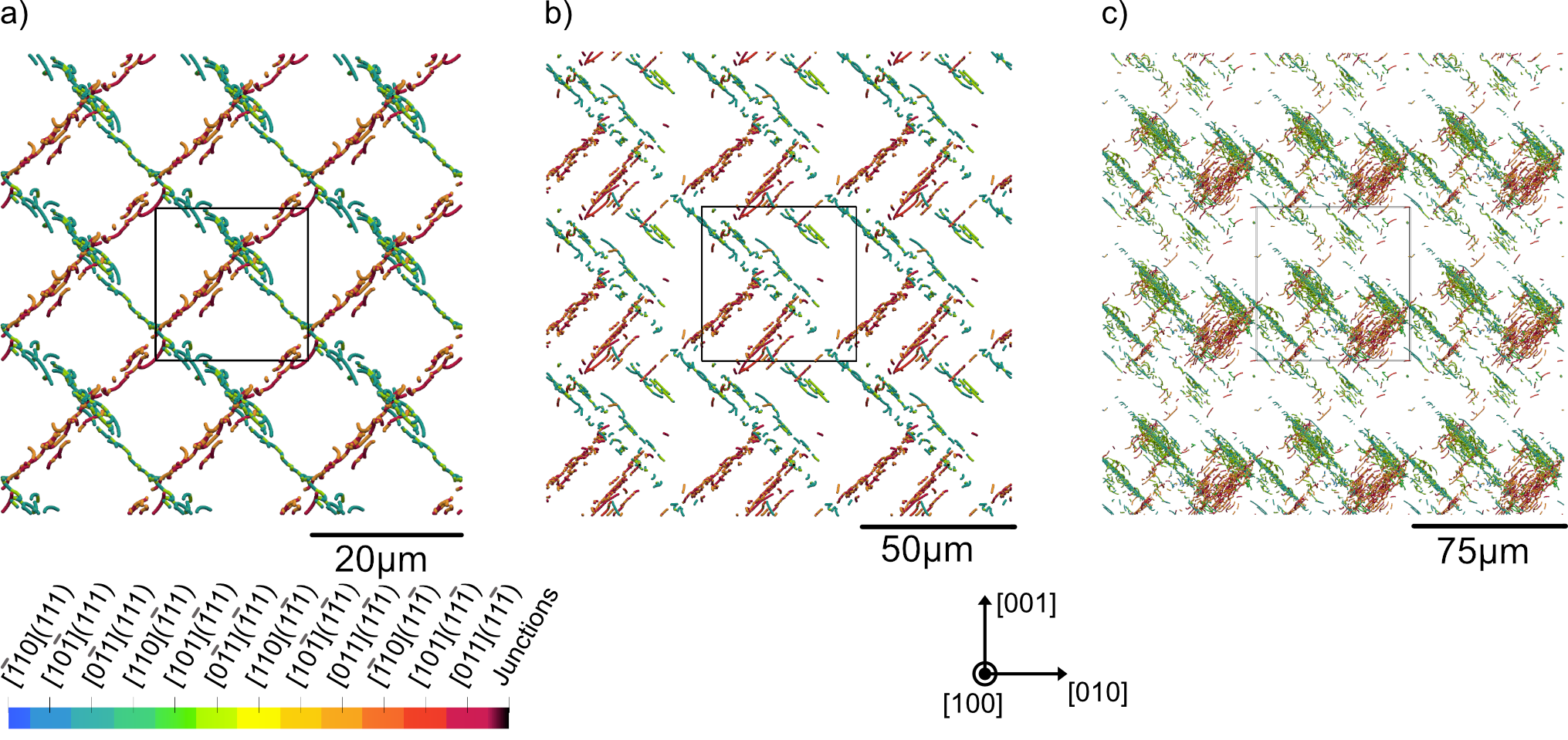}
    \caption{The predicted dislocation microstructure in a 1 $\mu$m thick slab with plane normal to the [100]-direction, which was extracted from the center of the 3D DDD simulation volumes shown in Fig.\,\ref{fig:appendix_Dislocation_microstructure_all_3d} for the: (a) $20^3\,\upmu$m$^3$; (b) $50^3\,\upmu$m$^3$; and (c) \seventy simulation cells. 
    The black lines represent the original simulation cell.}
    \label{fig:appendix_Dislocation_microstructure_full}
\end{figure}

\begin{figure}[h!]
    \centering
    \includegraphics[width=\textwidth]{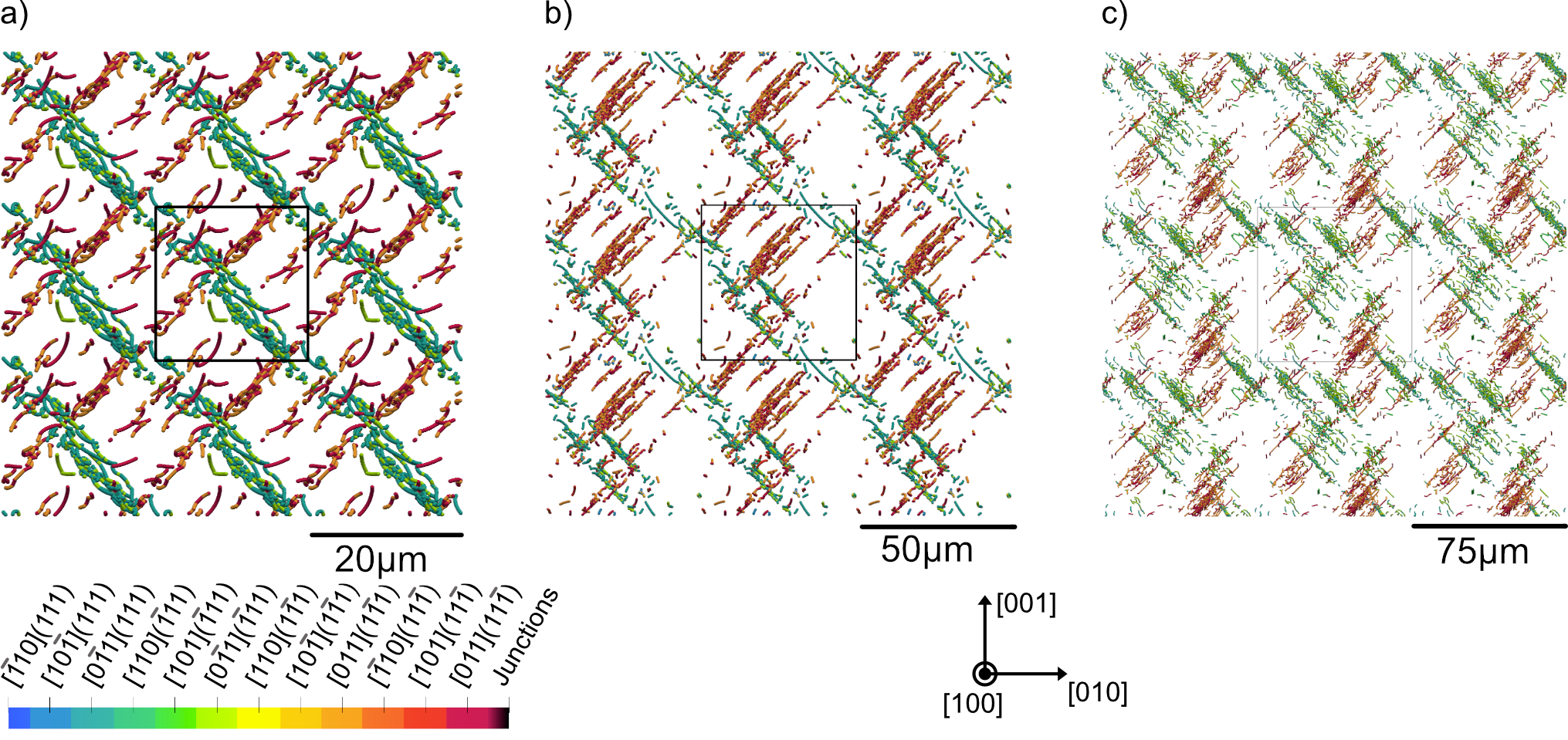}
    \caption{The predicted dislocation microstructure in a 1 $\mu$m thick slab with plane normal to the [100]-direction, which was extracted from the upper half of the 3D DDD simulation volumes shown in Fig.\,\ref{fig:appendix_Dislocation_microstructure_all_3d} for the: (a) $20^3\,\upmu$m$^3$; (b) $50^3\,\upmu$m$^3$; and (c) \seventy simulation cells. 
    The black lines represent the original simulation cell. The slabs where extracted at heights of: (a) $X \approx 6.5\,\mu$m; (b) $X \approx 14\,\mu$m; and (c) $X \approx 17.5\,\mu$m.}
    \label{fig:appendix_Dislocation_microstructure_full_Top}
\end{figure}

\begin{figure}[h!]
    \centering
    \includegraphics[width=\textwidth]{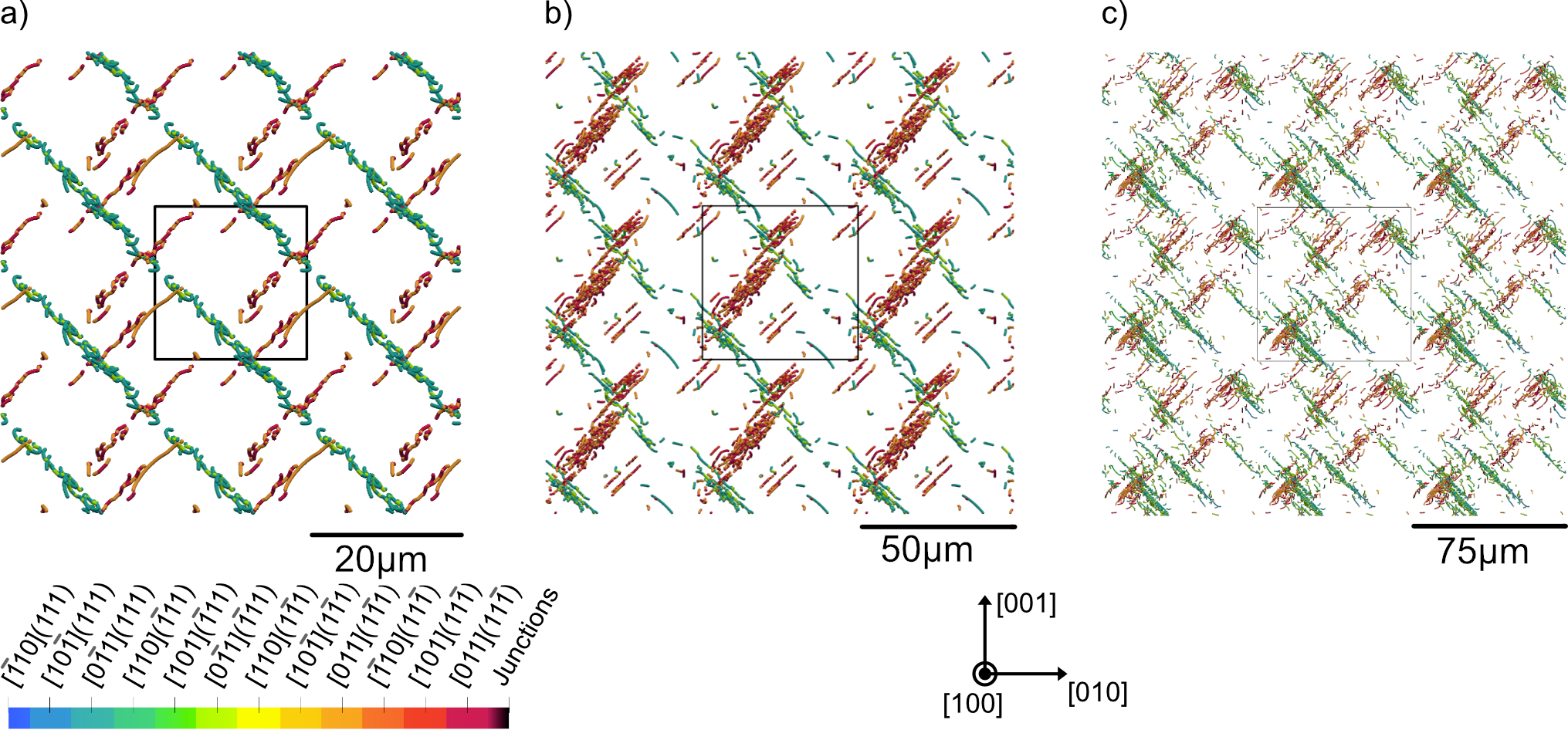}
    \caption{The predicted dislocation microstructure in a 1 $\mu$m thick slab with plane normal to the [100]-direction, which was extracted from the lower half of the 3D DDD simulation volumes shown in Fig.\,\ref{fig:appendix_Dislocation_microstructure_all_3d} for the: (a) $20^3\,\upmu$m$^3$; (b) $50^3\,\upmu$m$^3$; and (c) \seventy simulation cells. 
    The black lines represent the original simulation cell. The slabs where extracted at heights of: (a) $X \approx -6.5\,\mu$m; (b) $X \approx -14\,\mu$m; and (c) $X \approx -17.5\,\mu$m.}
    \label{fig:appendix_Dislocation_microstructure_full_Bottom}
\end{figure}

Furthermore, dislocation structures in $1\,\mu$m thick sections with plane normal in the $[100]$-direction  extracted from the 3D structures shown in Fig.\,\ref{fig:appendix_Dislocation_microstructure_all_3d} at from the center, top, and lower half of the simulation cell are shown in Fig.\,\ref{fig:appendix_Dislocation_microstructure_full}, Fig.\,\ref{fig:appendix_Dislocation_microstructure_full_Top}, and Fig.\,\ref{fig:appendix_Dislocation_microstructure_full_Bottom}, respectively.
It is observed that while the spacing of characteristic microstructure features (i.e., the spacing of dislocation walls) remain relatively constant throughout the height of the volume in the \twenty simulation cell case, the dislocation structure shows more randomness in case of the \seventy simulation cell. 
Although some periodicity is observed in case of the \fifty simulation cell, more heterogeneity in the dislocation structure is evident throughout the height of the volume as compared with the \twenty simulation cell. 
This indicates that the dislocation structure predicted by DDD simulations using the \fifty simulation cell is in transition to size-invariant periodicity.

\FloatBarrier

\renewcommand\refname{Supplemental References}

\bibliographystyle{unsrt}